\documentclass[preprint]{aastex}
\usepackage{graphicx}

\input{epsf.sty}
\input{psfig.sty}

\def\hb{H$\beta$ }
\def\hbe{H$\beta$}
\def\ha{H$\alpha$ }
\def\hae{H$\alpha$}
\newcommand{\nii}{[N\,II]}
\newcommand{\sii}{[S\,II]}

\def\mub{$\mu_0^{\rm B}$ }
\def\mube{$\mu_0^{\rm B}$}
\def\umu{mag\,arcsec$^{-2}$ }
\def\umue{mag\,arcsec$^{-2}$}

\def\feii{\ion{Fe}{2} }

\def\oiii{[\ion{O}{3}] }
\def\oiiie{[\ion{O}{3}]}

\def\kmps{${\rm km\,s^{-1}}$ }
\def\kmpse{${\rm km\,s^{-1}}$}

\begin{document}
\title{A Study of Active Galactic Nuclei in Low Surface Brightness Galaxies with Sloan Digital Sky Survey Spectroscopy}
\author{Lin Mei\altaffilmark{1,2} ,Weimin Yuan\altaffilmark{3} and Xiaobo Dong\altaffilmark{4} }
\date{Accepted ..... Received ......;}
\altaffiltext{1}{National Astronomical Observatories, Chinese Academy of Sciences, 20A Datun Road, Chaoyang District, Beijing 100020, China. E-mail: meilin05@mails.gucas.ac.cn}
\altaffiltext{2}{Graduate School of the Chinese Academy of Sciences, 19A Yuquan Road, P.O.Box 3908, Beijing 100039, China}
\altaffiltext{3}{Yunnan Observatory, National Astronomical Observatories, Chinese Academy of Sciences, Phoenix Hill, Kunming, Yunnan Province 650011, China. E-mail: wmy@ynao.ac.cn}
\altaffiltext{4}{Center for Astrophysics, University of Science and Technology of China, Hefei, 230026, China. E-mail: xbdong@ustc.edu.cn}

\begin{abstract}
Active galactic nuclei (AGN)
in low surface brightness galaxies (LSBGs) have received little attention
in previous studies. In this paper,
we present detailed spectral analysis of 194
LSBGs from the Impey et al. (1996) APM LSBG sample which have
been observed spectroscopically by the Sloan Digital Sky Survey Data Release 5 (SDSS DR5).
Our elaborate spectral analysis enables us to carry out, for the first time,
reliable spectral classification of nuclear activities in LSBGs
based on the standard emission line diagnostic diagrams in a rigorous way.
Star-forming galaxies are common, as found in about 52\% LSBGs.
We find, contrary to some of the previous claims,
that the fraction of galaxies containing an AGN
is significantly lower than that found in nearby normal galaxies of
high surface brightness.
This is qualitatively in line with the finding of Impey et al. (2001).
This result holds true even within each morphological type from Sa to Sc.
LSBGs having larger central stellar velocity dispersions, or
larger physical sizes, tend to have a higher chance to harbor an AGN.
For three AGNs with broad emission lines, the black hole masses
estimated from the emission lines are broadly consistent with
the well known M-$\sigma_\ast$
relation established for normal galaxies and AGNs.
\end{abstract}

\section{Introduction}

Galaxies with blue central surface brightness significantly fainter
than the classical Freeman value of \mube=21.65\,\umu are  
 commonly referred to as Low Surface Brightness Galaxies (LSBGs).
The exact defining criterion of \mube,  which is the 
central surface brightness of a galactic disk by convention, 
varies in the literature, though it falls mostly within 
\mube=22.0 --23.0\,\umu \citep{oneil98,bell}. 
It has been suggested that LSBGs
could account for a significant fraction 
of all galaxies in the Universe \citep{b53,b16},
and thus are an important constituent of galaxies.
LSBGs show some extremely different properties from high surface brightness galaxies (HSBGs).
The typical value of metallicity in LSBGs is 
one third of the solar metallicity \citep{impey97}.
Observationally, like Malin 1 \citep{b13},
a significant number of LSBGs possess diffuse faint disks with little stellar
content but substantial amounts of neutral hydrogen gas.
The low surface brightnesses indicates low star formation rates (SFRs) in these systems.
Indeed, it has been found that the HI surface mass densities in LSBGs are near
or below the critical gas surface density for star formation threshold \citep{b14,b23,b5,b19}.
These extreme properties imply that LSBGs are less evolved compared to HSBGs.
Bulge-dominated LSBGs are redder than disk-dominated LSBGs and both can
be well described as exponential surface brightness distribution \citep{b1}.
Bulges of LSBGs were found to be metal-poor compared to those hosted by HSBGs \citep{b7}.
\citet{b20} found that there is no evidence for heavy dust obscuration in LSBGs.
However, a recent study of infrared properties of LSBGs using Spitzer data indicated
that modest amounts of dust exist in a fraction of LSBGs, 
although their metallicity and apparent transparency are low \citep{b10}.

The low  surface brightness, 
comparable to or fainter than 
the night sky background (22.5 $-$ 23.0 mag arcsec$^{-2}$ in the B band),
makes the detection of LSBGs difficult.
As such, much less optically selected LSBGs samples than normal HSBGs have been cataloged 
from surveys by ground-based telescopes (e.g., Impey et al. 1996; Monnier Ragaigne et al. 2003).
\citet{b12} published a LSBG catalog of 693 LSBGs derived from the
Automated Plate Measuring (APM) machine scans of
the UK Schmidt Telescope survey plates.
A catalog containing 2469 southern-sky LSBGs also from the APM scans
of the UK Schmidt photographic plates was given by
\citet{morshidi}.

Unlike HSBGs, which have been the focus of extensive  studies in
extragalactic astrophysics over many years, nuclear activity in
LSBGs has drawn little attention. This may be due partly to lack
of large samples of LSBGs. So far, few secured AGNs with reliable
detections have been reported in the literature, not mentioning
their properties. Potentially, a study of AGNs in
LSBGs is as important as in HSBGs for the following reasons. First,
it may provide important and complementary  clues to uncover the
formation and growth of super-massive black holes (SMBH), since
LSBGs may have experienced a different route from HSBGs in their
formation and evolution. Second, in light of AGN feedback, the
evolution and ecology of LSBGs may be affected by the presence of
powerful AGNs. Third, a comparison of different host galaxies
environments where AGNs reside in, HSBGs and LSBGs,  may shed light on
the triggering of AGN activity and the dependence of AGN properties
on host galaxies.

However, few yet contradicting results have been given in the literature
as to the fraction of LSBGs harboring AGNs.
When AGN activity was first searched for via optical spectroscopy
in small samples of LSBGs,
a surprising result was suggested that LSBGs seem to have a higher
fraction of AGNs.
Among 10 giant LSBGs, Sprayberry et al. (1995) found
4 Seyfert\,1 and 1 Seyfert\,2 nuclei, indicating a significantly high
AGN fraction.
In a sample of 34 giant, HI-rich LSBGs,  Schombert (1998)
found a high fraction ($40-50$\%) of low luminosity AGNs, similar to that
in the magnitude-limited local galaxy sample studied by \citet{b11a},
which are mostly HSBGs.
However, such a high AGN fraction could not be confirmed by
\citet{b38} in a spectroscopic study of 250 LSBGs drawn from the
Impey et al. (1996) LSBG sample;
in contrast, they found an AGN fraction as low as less than 5$\%$.

It should be noted that in essentially all those previous studies,
the identification of AGNs was subjected to  large
uncertainties, due to the following limitations.
Firstly, the optical spectra taken were mostly of low spectral resolution
(a resolution $\sim20\AA$ used in Impey01)
and low signal to noise ratio (S/N),
and some of narrow wavelength coverage.
As such, the line flux measurements were uncertain, especially
those close lines suffering from blending.
In most studies above,
AGNs were identified as having strong low-ionization
features ([NII] and [SII]) combined with [OI],
unless a broad Balmer line was present, rather than based on
the rigorous line flux ratio diagnostic diagrams.
The only case where an emission line ratio diagnostic diagram
([OIII]/\hb -- [SII]/\hae)
was invoked to separate AGNs and HII regions was in
Impey01.
However, there are considerable uncertainties in the line ratios,
especially for \hae, which is heavily blended with [NII]$\lambda\lambda$6548,6583.
Secondly, no subtraction of starlight spectrum of the host galaxy
was performed in those studies,
which is important for search of AGN signatures,
especially for low luminosity AGNs, as discussed in detail in \citet{b11b}.
The spectra of host galaxy starlight
could affect  severely the detection and measurement of
emission lines, by making some weak emission line invisible (such as \hbe),
or distort the line fluxes,
or even spuriously mimic weak broad emission lines.

The Sloan Digital Sky Survey (SDSS, \citet{b26})
has acquired high quality optical
spectra of a million galaxies over a large portion of the sky \citep{b47},
some of which could be LSBGs.
On the other hand, we have developed an algorithm for successful removal of
host galaxy starlight and for accurate emission line spectral fit
(see Zhou et al. 2006 for details).
These two provide us with an opportunity
to revisit the above unsolved question of AGNs in LSBGs,
with much better spectral S/N and
resolution ($R=1800-2200$), wavelength coverage, and data homogeneity.
The high spectral quality of large LSBG samples provided by the SDSS
allows us  to study not only the AGN demographics, but also,
for the first time, the properties of AGNs in LSBGs.
This paper presents such a study of nuclear activity in a LSBG sample.
We introduce the sample and the data analysis in Section 2.
The demographics of nuclear activity in LSBGs are presented
in Section\,3, followed by a preliminary study of the properties of LSBG AGNs
in Section\,4. Discussion and a summary of conclusions are given in
the last two sections, respectively.
A cosmology with H$_{0}$ = 70 km s$^{-1}$ Mpc$^{-1}$, $\Omega_M$ = 0.3,
and $\Omega_\Lambda$ = 0.7 is adopted.

\section[]{APM--SDSS sample and spectral data analysis}
\subsection{The APM-SDSS sample}

In the northern sky, 
the largest, well-defined optically selected LSBG sample is the
catalog compiled by Impey et al. (1996), which was constructed from
the Automated Plate Measuring (APM) machine scans of
the UK Schmidt Telescope survey plates.
It is the most extensive catalog of LSBGs in the northern sky to date,
comprising 693 galaxies in the local universe with redshifts $<0.1$
selected from a sky region of 786 square degrees centered on the equator.
Among them,  513 galaxies have
large angular sizes (The major-axis diameter D $\geq$ $30\arcsec$ at
the limiting isophote of the APM scans of 26mag arcsec$^{-2}$) and
180 have small angular sizes (D $<$ $30\arcsec$).
Most LSBGs in this catalog have central surface brightness ranging from
\mube $=$21.5mag arcsec$^{-2}$ to 26.5mag arcsec$^{-2}$
in the B band.
It should be noted that some of the galaxies with \mub
at the bright end may not qualify as LSBG when the \mube=22.0\,\umu
cutoff is applied.
However, a significant fraction of these galaxies 
is still expected to be LSBGs, since the presence
of a galactic bulge residing in a low surface brightness (LSB) disk
may apparently result in a bright \mube\ \citep{b1}.

\begin{figure}
\includegraphics[width=0.8\hsize,height=0.45\hsize]{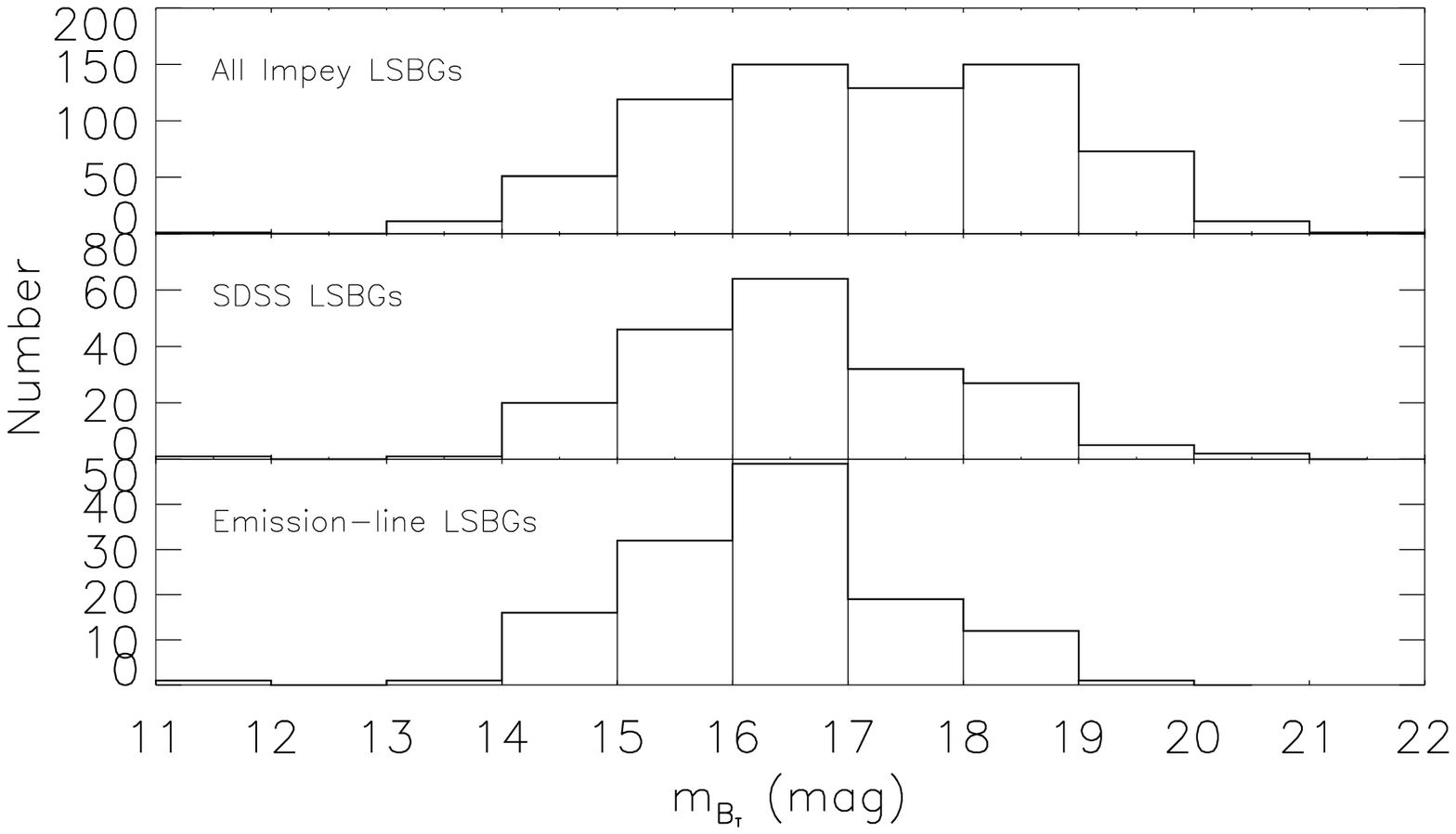}
\includegraphics[width=0.8\hsize,height=0.45\hsize]{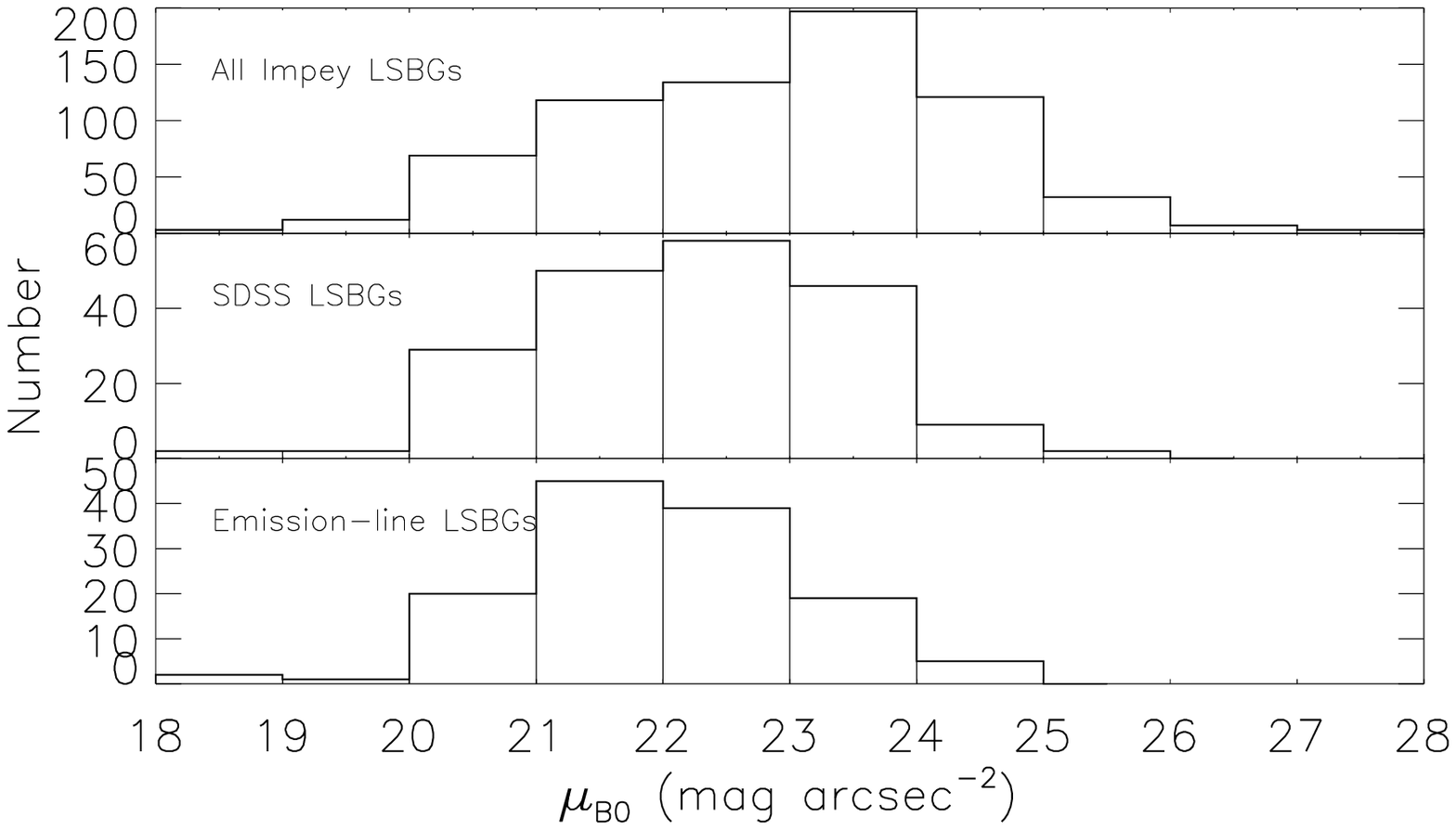}
\caption{Distributions of apparent magnitude m$_{B_T}$ in the B band, 
central surface brightness \mube\ (top panel: the 693 LSBGs from the catalog of Impey et al. (1996); 
middle panel: the 194 APM-SDSS LSBGs; 
bottom panel: the 131 emission-line LSBGs with the H$\beta$, [OIII] $\lambda$5007, 
H$\alpha$ and [NII] $\lambda$6583 four lines detected with S/N $>$ 3).}
\label{lsbg_bt_mu0}
\end{figure}

We searched for spectral data  from the SDSS Data Release 5 (DR5)
for the LSBGs in the Impey et al. sample.
We found that, out of the 693 LSBGs, 194 have SDSS spectra
and were classified as galaxies by the SDSS pipeline.
They form our sample of study in this paper.
Among them, 95 are also in the spectroscopically observed subsample
presented by Impey01.
Figure\,\ref{lsbg_bt_mu0} shows the distributions of 
total $B$-magnitude and \mub 
for the whole APM sample and the APM-SDSS subsample.
It can be seen that the APM-SDSS subsample spans almost the
entire ranges of the total magnitude and \mube, 
though it drops more quickly at the faint end 
than the parent sample, due to the magnitude limit of the 
SDSS spectroscopic survey.
It also shows that a large fraction of the  APM-SDSS subsample
has \mub fainter than $22$\,\umue, the minimum of the nominal  
threshold for LSBGs.
For the other objects with  \mube$<22$\,\umue, 
some are likely LSBGs with a (dominating) bulge.
Therefore the APM-SDSS subsample can be regarded as being composed of
mostly LSBGs, with possible inclusion of some intermediate galaxies
in between the typical LSBG and HSBG types.
For simplicity, we refer to all the sample objects as LSBGs nominally.

\subsection{Spectral data analysis}

The SDSS spectra have a wavelength coverage from 3800 to 9200 \AA\
with a resolution of R$\sim$1800--2200.
As was demonstrated by \citet{b11a},
the vast majority of the AGN population in the local universe 
is low luminosity AGN (LLAGN).
For LLAGNs, the optical spectra 
taken through either slits or fibers are dominated by host galaxy starlight.
The SDSS spectra were taken through 
a fiber aperture of 3\arcsec\ in diameter (corresponding to 2.9 kpc at a redshift of 0.05),
and thus include large amounts of starlight from the host galaxies
for the APM-SDSS sample objects.
Careful removal of starlight, especially stellar absorption features,
is essential for the detection and measurement of 
possible nuclear emission lines.
For proper modelling of host galaxy starlight spectra,
we use the algorithm developed at Center for Astrophysics, 
University of Science and Technology of China,
which is described in detail in Zhou et al. (2006, see also Lu et al. 2006)
and is summarized in Appendix\,A in this paper.
 
After subtracting modeled stellar spectra and a power-law continuum, 
the leftover emission line spectra, if any, are fitted
using an updated version of the code described in \citet{dong05},
with improvement made for better recovery of weak emission lines.
The broad \ha\ and \hb\ lines, if present, are hard to separate
from nearby narrow lines.
As the narrow Balmer lines and the \nii$\lambda\lambda$6548, 6583 doublet
have similar profiles to the \sii$\lambda\lambda$6716, 6731 doublet
\citep{Filippenko88,b11b,b36},
we use the \sii$\lambda\lambda$6716, 6731 doublet lines as a template
to fit narrow lines.
If \sii\ is weak, \oiii$\lambda$5007 is then used as the template.
The \sii\ doublet are assumed to have the same
profiles and redshifts;
and each is fitted with as many Gaussians as is statistically justified,
generally with 1--2 Gaussians.
Likewise, the \oiii$\lambda\lambda$4959, 5007 doublet are fitted in a similar way.
Furthermore, the flux ratio of $\lambda$5007/$\lambda$4959 is fixed to
the theoretical value of 3.
When a good model of the narrow-line template is achieved,
we scale it to fit the
narrow Balmer lines and the \nii$\lambda\lambda$6548, 6583 doublet.
The flux ratio of the \nii\ doublet $\lambda$6583/$\lambda$6548 is fixed to
the theoretical value of 2.96.
For the possible broad \ha\ and \hb\ lines,
we use multiple Gaussians to fit them, as many as is statistically justified.
If a broad emission line is detected at $\geq 5 \sigma$ confidence level,
we regard it as genuine.
If the broad \hb\ line is too weak to achieve a reliable fit,
we then re-fit it assuming that it has the same profile and redshift as the broad \ha\ line.

\begin{figure}
\includegraphics[width=1.\hsize,height=0.4\hsize]{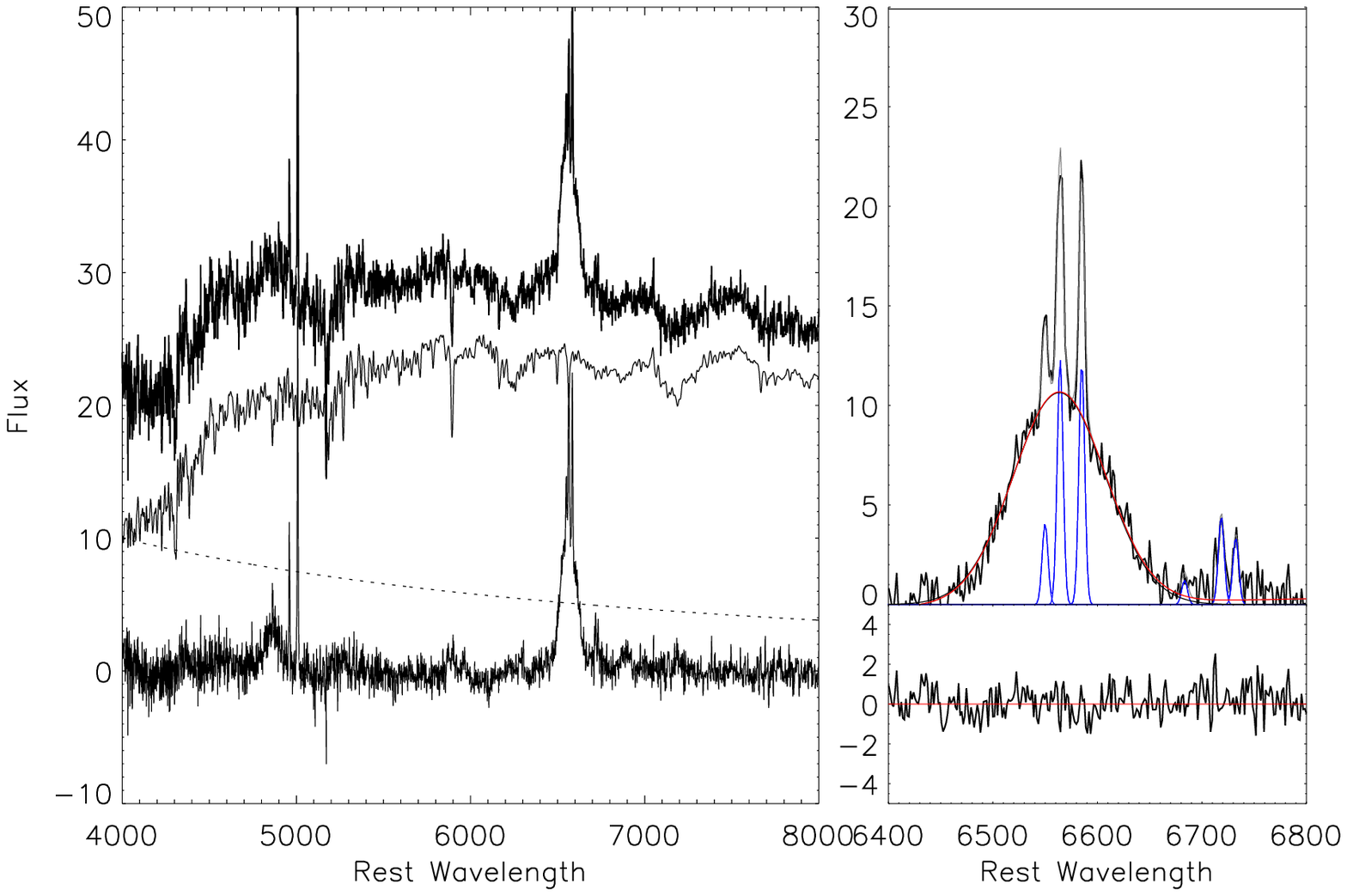}
\includegraphics[width=1.\hsize,height=0.4\hsize]{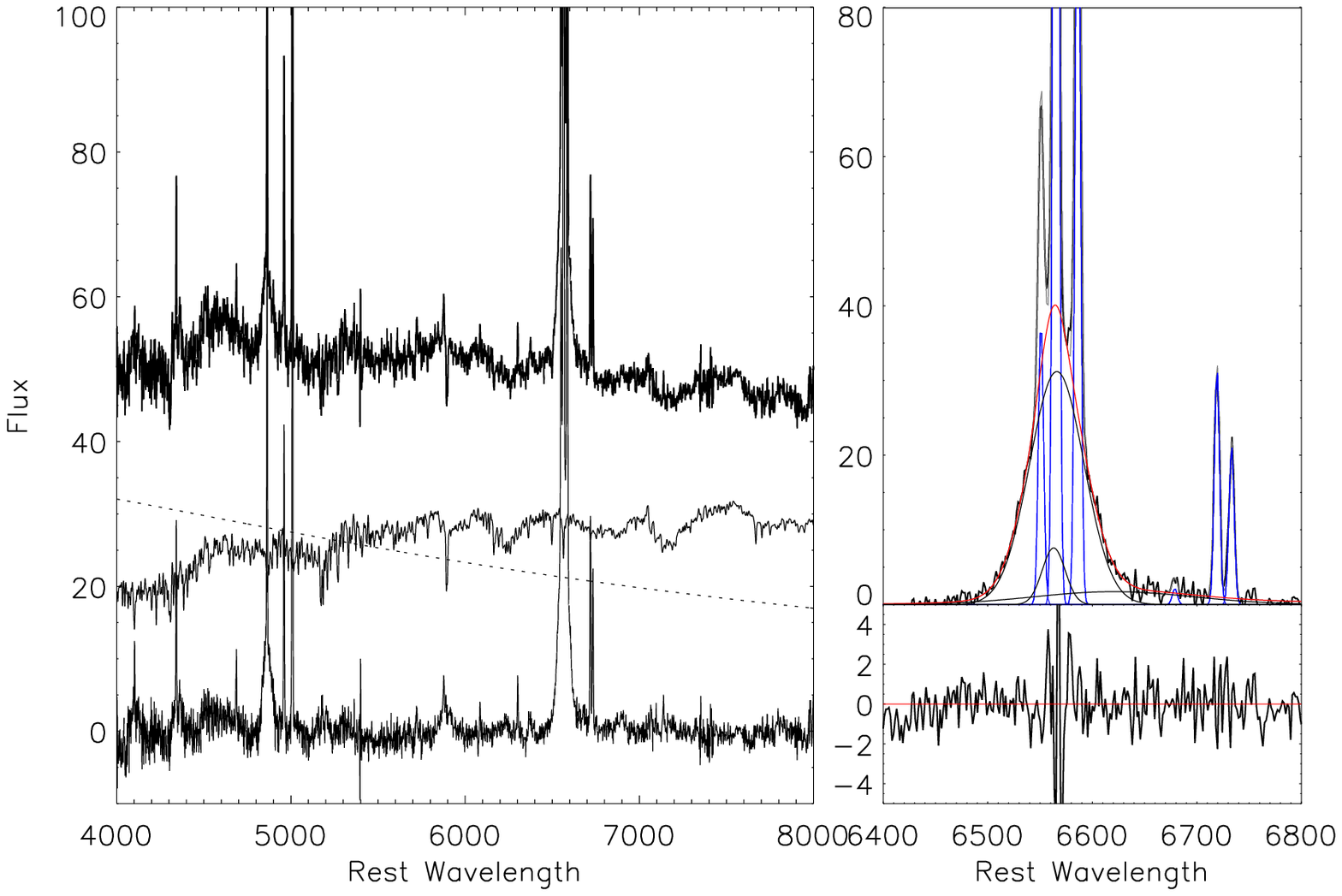}
\includegraphics[width=1.\hsize,height=0.4\hsize]{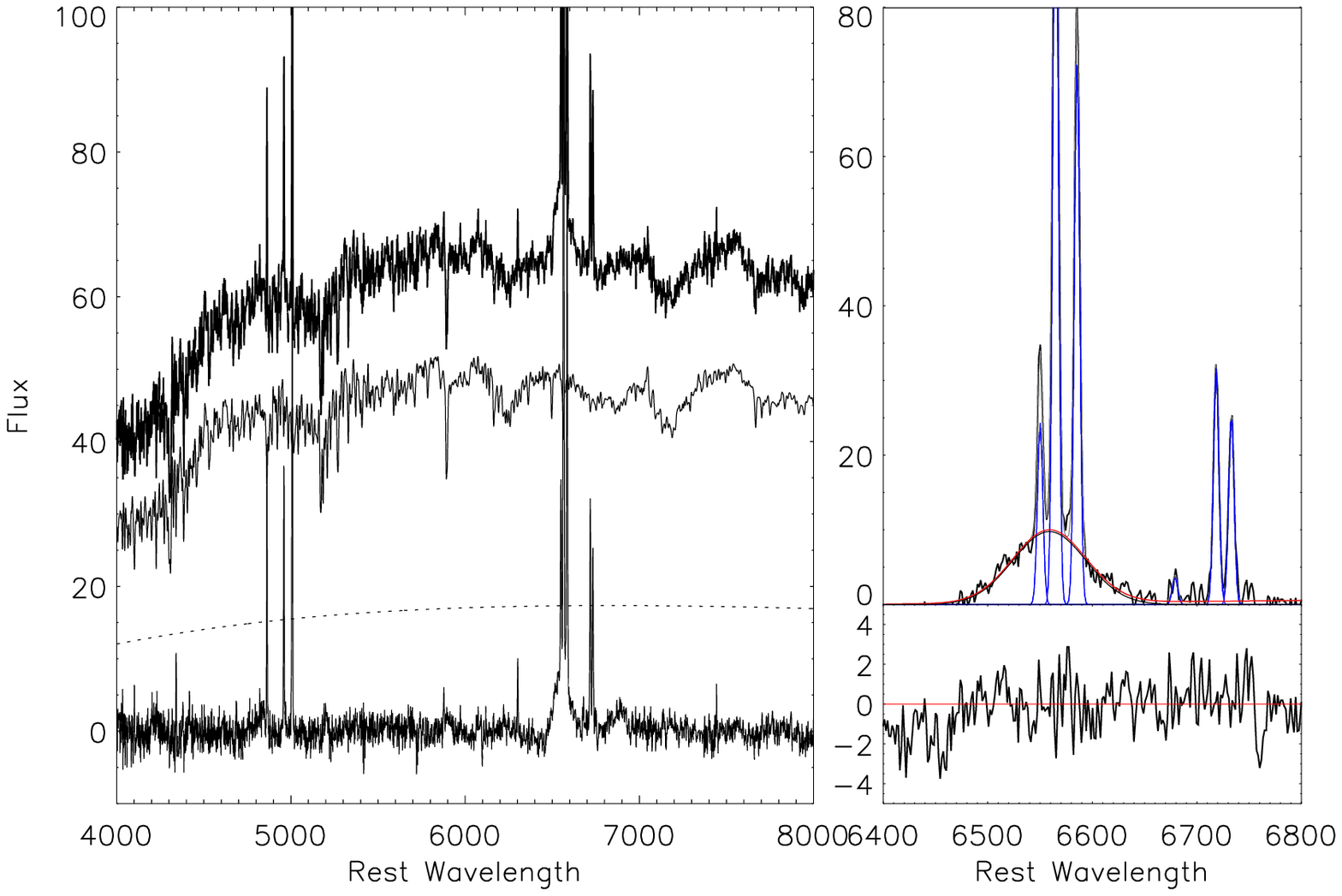}
\caption{SDSS spectra of 3 LSB AGNs showing broad lines.
The vertical axis is flux in unit of 10$^{-17}$
erg~s$^{-1}$~cm$^{-2}$~\AA $^{-1}$ and the horizontal is
wavelength in \AA . The left panel shows the procedure of
starlight/continuum subtraction: from top to bottom, the original spectrum, the
stellar component, the nuclear continuum, and the starlight/continuum subtracted
residual. The right panel shows the result of spectral fit in the H$\alpha$-[SII] region 
(Top panel: original data and individual components of the fit. 
Blue curve: narrow component; 
red curve: Broad component; 
gray curve: final fit result.
Lower panel: the residuals). 
Top: SDSS J122912.9$+$004903.7; 
middle: SDSS J011448.7$-$002946.1; 
bottom: SDSS J231815.7$+$001540.2.}
\label{fit_broad_lsbg}
\end{figure}

Our analysis results in detections of 131 
emission line galaxies out of the total 194 LSBGs.
This gives a fraction of 68\% of emission line galaxies in LSBGs.
The objects are list in Table\,\ref{131_lsbgs}, along with the
fitted parameters of the important lines. 
Also listed are the stellar velocity dispersions in the central
region of the host galaxies, $\sigma_*$, which are obtained in the
above procedure of modeling the host galaxy starlight spectra.
Three out of the 131 galaxies are found to show broad Balmer 
emission lines and should be broad line AGNs.
Their spectra are shown in Figure\,\ref{fit_broad_lsbg}
and the parameters of the broad lines are given in Table\,\ref{3_broad_lsbgs}.
It should be noted that the widths of emission line are corrected for
the instrumental broadening, which is 55-80 km s$^{-1}$ for SDSS spectra.

\begin{figure}
\includegraphics[]{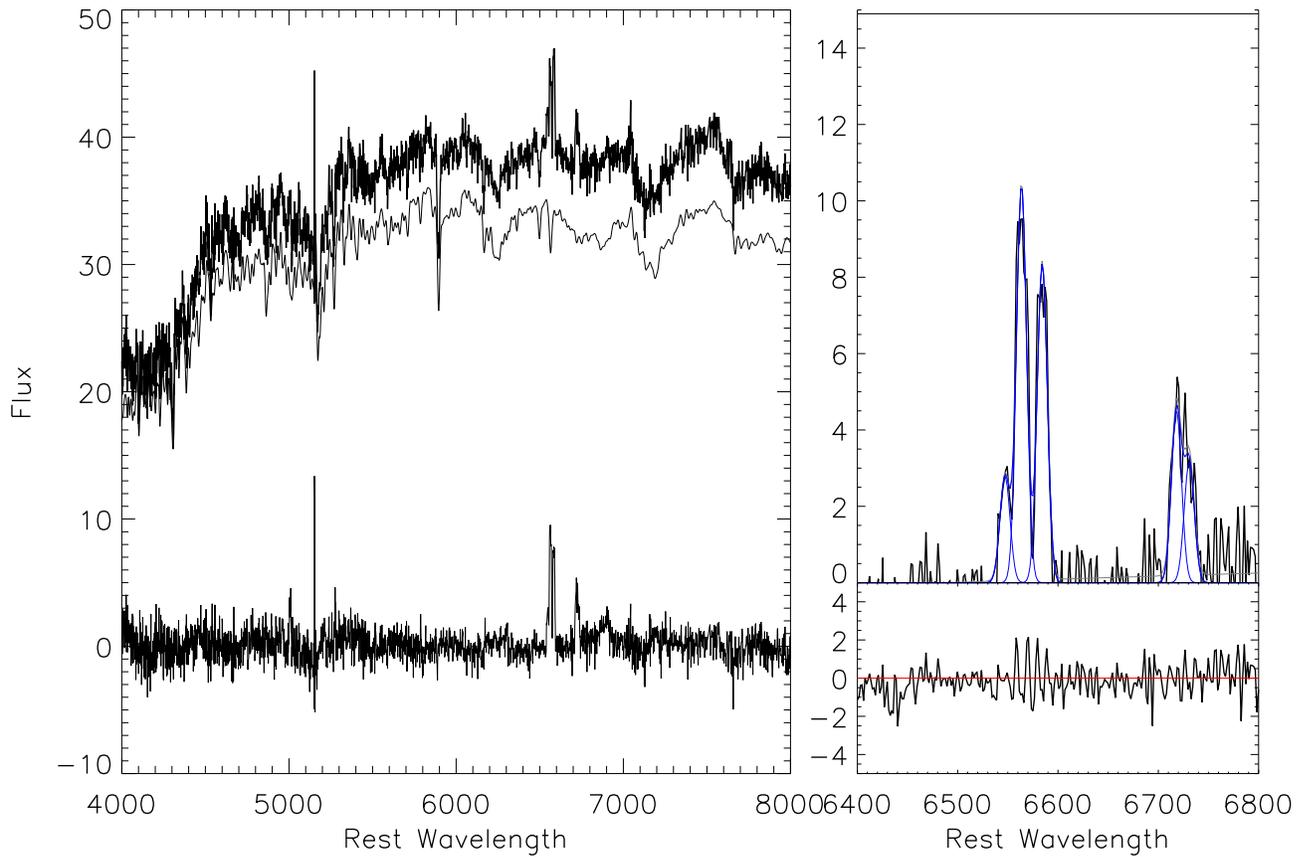}
\caption{The same as Figure.\ref{fit_broad_lsbg}. 
SDSS spectrum of the LSBG 1436$+$0119 
(SDSS J143846.3$+$010657.7) which was claimed to have a broad H$\alpha$ line by Impey01. 
No broad H$\alpha$ line is shown in the SDSS spectrum.}
\label{impey_broad}
\end{figure}

To demonstrate the reliability of our spectral analysis,
as well as the spectral quality of the SDSS,
we compare our detections of broad line AGNs with those 
in Impey01.
Among the 95 overlapping objects in both the Impey01
spectroscopic subsample and our APM-SDSS subsample,
two were  claimed to have broad emission lines by the authors.
1226$+$0105 (SDSS\,J122912.9$+$004903.7) is also found in our work
with the full width at half maximum (FWHM) of \ha 4750\,km\,s$^{-1}$, which is slightly broader
than the previous value (4570\,km\,s$^{-1}$, Impey01).
The narrower FWHM value in Impey01 is likely due to 
the fact that broad and narrow component deblending
could not be performed given the low spectral resolution (20$\AA$).
For the second object, 1436$+$0119 (SDSS J143846.3$+$010657.7), however, 
we cannot confirm the presence of a broad \ha line 
with the starlight subtracted SDSS spectrum (Figure\,\ref{impey_broad}).
An inspection of the original spectrum in Impey01
seems to suggest that the
previously claimed broad \ha line is likely spurious and
just a feature in the stellar spectrum
(no starlight subtraction was performed in Impey01)
when the resolution and S/N is low.
Furthermore, the one with the weakest broad line 
among the above 3 broad line AGNs in our work 
(see Figure\,\ref{fit_broad_lsbg}),
SDSS J231815.7$+$001540.2 (NGC\,7589),  
was missed in Impey01.
This object is likely a Seyfert\,1.9, which was previously 
found to show strong, highly variable hard X-ray emission \citep{b24}.
We thus conclude that our spectral analysis results are 
much more accurate and reliable.

It is worth noting that 
the three LSBGs, J005342.7$-$010506.6, 
J111549.4$+$005137.5 and J133032.0$-$003613.5 that were
classified as type 1 AGNs by \citet{b56}, 
do not show evident broad emission lines in this work.
This is likely due to the difference in subtraction of narrow lines.
\citet{b56} modeled every narrow line with one single Gaussian.
Yet we noticed that the profiles of narrow lines are asymmetric
and thus we used 2 Gaussians to model each of the [SII] doublet;
the fit is used as a model to subtract the narrow component of H$\alpha$ and  
[NII], as described in section 2.2.
The asymmetric profiles of the narrow H$\alpha$ and [NII] components
can mimic a broad H$\alpha$ component 
(see also Ho et al. 1997b; Greene \& Ho 2007).
To be conservative, we do not consider these 
3 LSBGs as broad line AGNs in this paper.

\subsection{Spectral classification of nuclear Activity}

Generally, there are three types of emission line spectra in galactic nuclei:
star-forming (HII) nuclei in which line emitting gas is  photoionized by 
radiation from hot stars, 
Seyfert nuclei characteristic of  photoionization by a power-law continuum
powered by black hole accretion,
and Low-Ionization Nuclear Emission Regions (LINERs, Heckman 1980)
with relatively strong low ionization lines 
(such as [OI] $\lambda$6300 and [NII] $\lambda\lambda$6548,6583)
and generally lower nuclear luminosities
compared to Seyfert galaxies.
The  nature of LINER is still controversial, 
though an accretion-powered AGN
appears to be preferred in recent studies \citep[see, e.g.][]{b11a}.
Following Ho et al., we regard LINERs as a subclass of AGN in this paper.

Classification of galactic nuclei based on optical emission lines
has been extensively investigated by various authors.
One common and effective method is to compare the flux ratios of 
close lines with predictions of different photoionization models.
It was first shown by \citet{b40} that AGNs could be separated from HII nucleus 
on the diagrams of 
the line flux ratios
[NII]$\lambda$6583 / H$\alpha$ vs. [OIII]$\lambda$5007 / H$\beta$,
[OII]$\lambda$3727 / [OIII]$\lambda$5007 vs. 
[OIII]$\lambda$5007 / H$\beta$, and
[OI]$\lambda$6300 / H$\alpha$ vs.\ [OIII]$\lambda$5007 / H$\beta$ 
(the BPT diagrams). 
\citet{b41} revised this method by including the line ratio
[SII]$\lambda\lambda$6717,6731/H$\alpha$
and excluding the [OII]$\lambda$3727/[OIII] $\lambda$5007 
because it is sensitive to reddening.
A similar scheme was also used by \citet{b11b}.
Using a large sample of SDSS emission line galaxies,
\citet{b44} found a clear, empirical separation between
star-forming galaxies and AGNs on the BPT diagram of 
[NII]$\lambda$6583/H$\alpha$ versus [OIII]$\lambda$5007/H$\beta$.
This classification scheme was refined recently by  \citet{b43}, by
including new criteria to separate Seyferts from LINERs using
the [SII]$\lambda\lambda$6717,6731/H$\alpha$
and [OI]$\lambda$6300/H$\alpha$ line ratios.
In this scheme, galaxies falling in between 
the empirical star-forming and AGN dividing line of Ka03 and 
the theoretical limit (maximum star-formation) of \citet{b42}
are regarded as composite objects 
(also referred to as transition objects),
which likely contain a metal-rich stellar population 
plus an AGN, either a Seyfert or a LINER (Ho et al. 1997a; Ke06)
In this paper, we adopt the classification scheme of Ke06, 
since it is the most recent updates and also easy to use.
Our careful deblending of the [NII] and [SII] doublets 
ensure that our spectral classification based on these criteria should be reliable.

The  classification scheme of Ke06 is summarized here.
First, star-forming galaxies are separated from AGNs using the
 Ka03 dividing line in the
[NII]$\lambda$6583/H$\alpha$ versus [OIII]$\lambda$5007/H$\beta$ diagram:
  star-forming galaxies are those falling 
below and to the left-hand side of the dividing line, i.e.
\begin{equation}
  \log([OIII] / H\beta) > 0.61 / \{\log([NII] / H\alpha) - 0.05\} + 1.3,
\end{equation}
and  AGNs are those falling above and to the right-hand side of 
the dividing line, i.e.\
\begin{equation}
  \log([OIII] / H\beta) < 0.61 / \{\log([NII] / H\alpha) - 0.05\} + 1.3.
\end{equation}

AGNs are further grouped into 3  subclasses:
composite galaxies are those falling in between the Ka03 and Ke01 
dividing lines in the
[NII]$\lambda$6583/H$\alpha$--[OIII]$\lambda$5007/H$\beta$ diagram:
\begin{equation}
  \log([OIII] / H\beta) > 0.61 / \{\log([NII] / H\alpha) - 0.05\} + 1.3
\end{equation}
and
\begin{equation}
  \log([OIII] / H\beta) < 0.61 / \{\log([NII] / H\alpha) - 0.47\} + 1.19
\end{equation}
Objects falling above the Ke01 maximum starformation line
are Seyfert galaxies or LINERs, i.e.\
\begin{equation}
  \log([OIII] / H\beta) > 0.61 / \{\log([NII] / H\alpha) - 0.47\} + 1.19.
\end{equation}
These two types are further separated in the diagrams of
[SII]$\lambda\lambda$6717,6731/H$\alpha$--[OIII] $\lambda$5007/H$\beta$
or [OI]$\lambda$6300/H$\alpha$--[OIII]$\lambda$5007/H$\beta$, i.e.
for Seyfert galaxies,
\begin{equation}
1.89 \log([SII] / H\alpha) + 0.76 < \log([OIII] / H\beta)
\end{equation}
or
\begin{equation}
1.18 \log([OI] / H\alpha) + 1.30 < \log([OIII] / H\beta);
\end{equation}
and for LINERs
\begin{equation}
1.89 \log([SII] / H\alpha) + 0.76 > \log([OIII] / H\beta)
\end{equation}
or
\begin{equation}
1.18 \log([OI] / H\alpha) + 1.30 > \log([OIII] / H\beta).
\end{equation}

\begin{figure}
\includegraphics[width=1.\hsize,height=0.4\hsize]{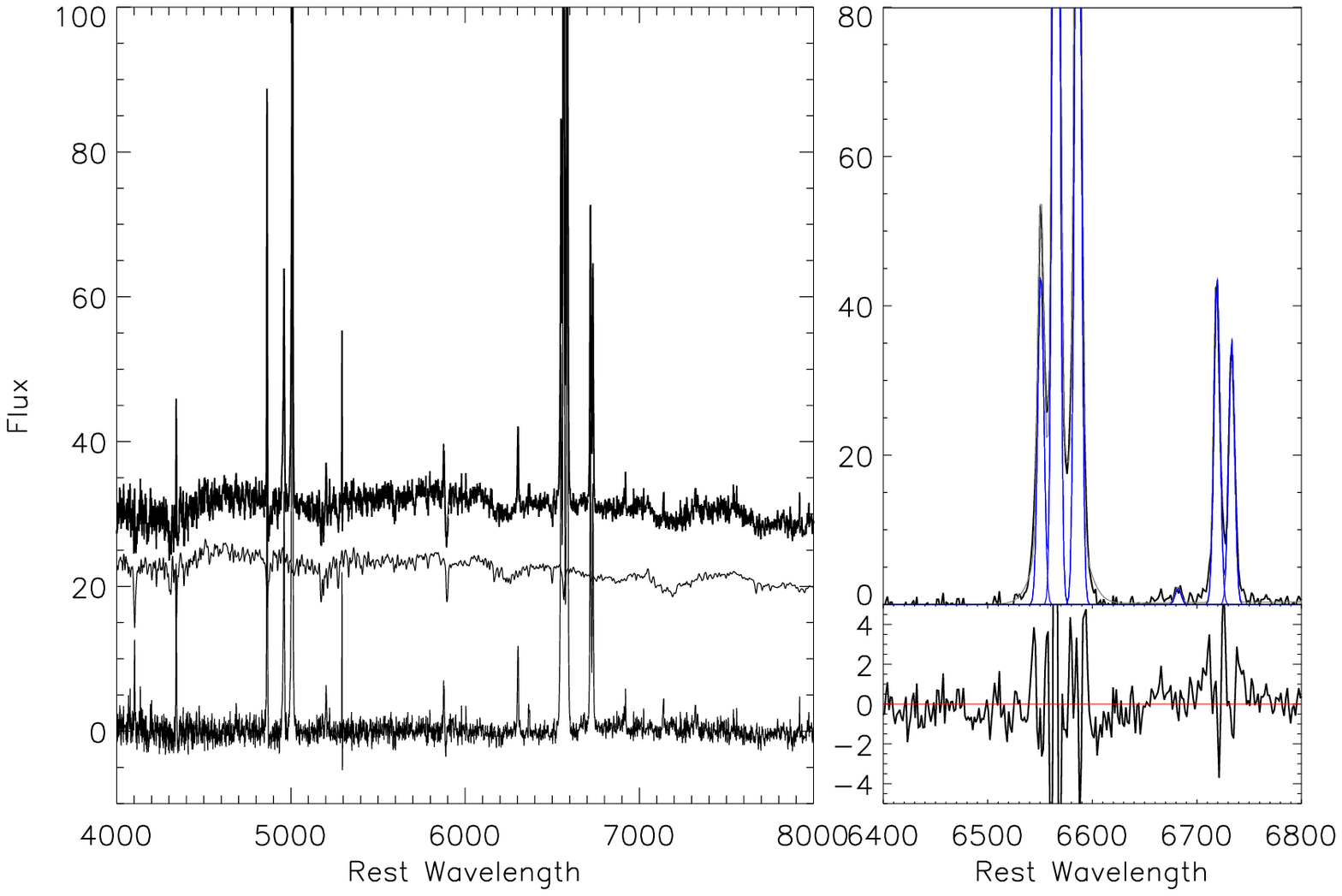}
\includegraphics[width=1.\hsize,height=0.4\hsize]{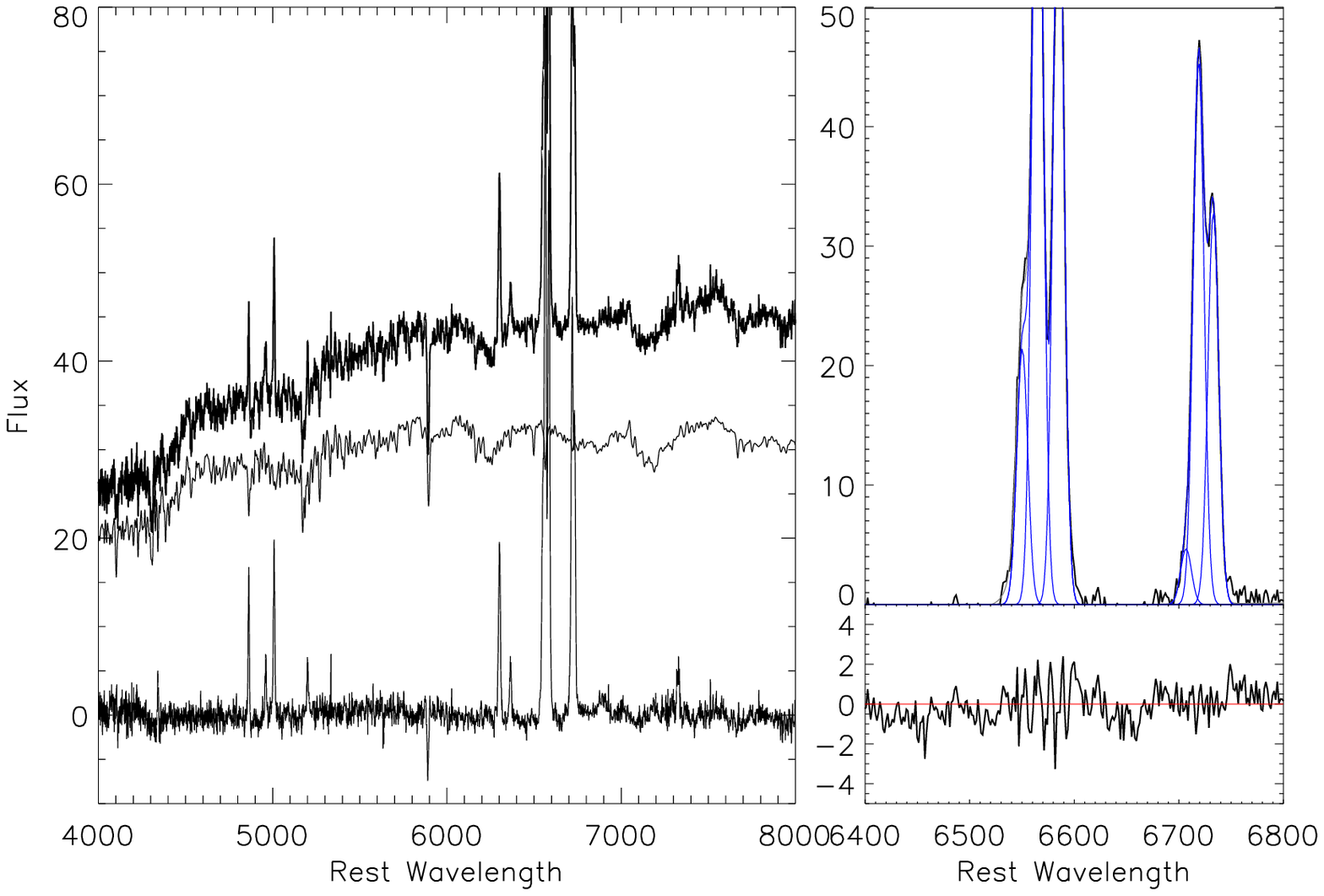}
\includegraphics[width=1.\hsize,height=0.4\hsize]{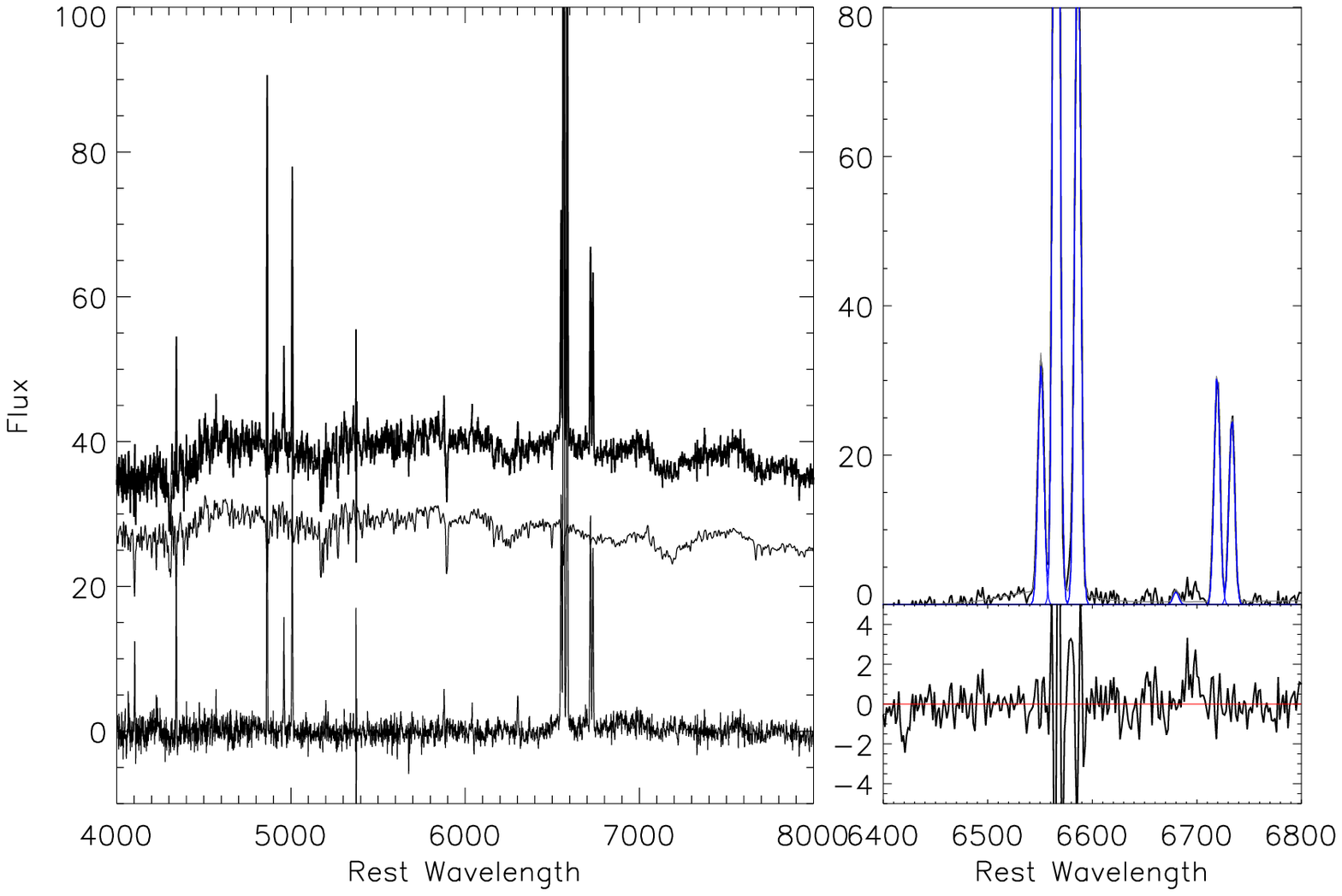}
\caption{The same as Figure.\ref{fit_broad_lsbg}. 
Demonstration spectra for the 3 types of nuclear activities. 
Top: Seyfert 2; 
middle: LINER; 
bottom: composite.}
\label{fit_narrow_lsbg}
\end{figure}

\begin{figure}
\includegraphics[width=0.8\hsize,height=0.4\hsize]{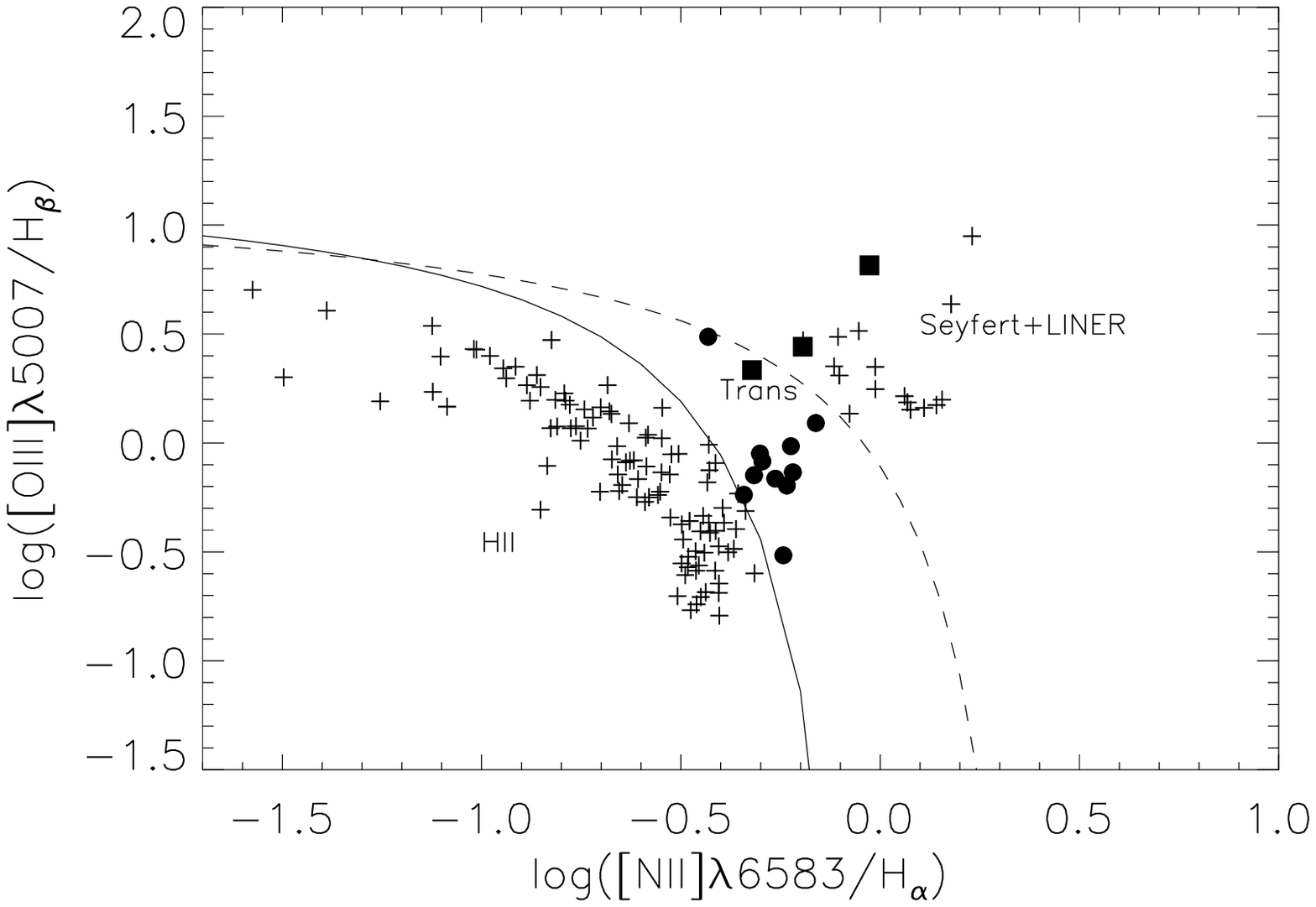}
\includegraphics[width=0.8\hsize,height=0.4\hsize]{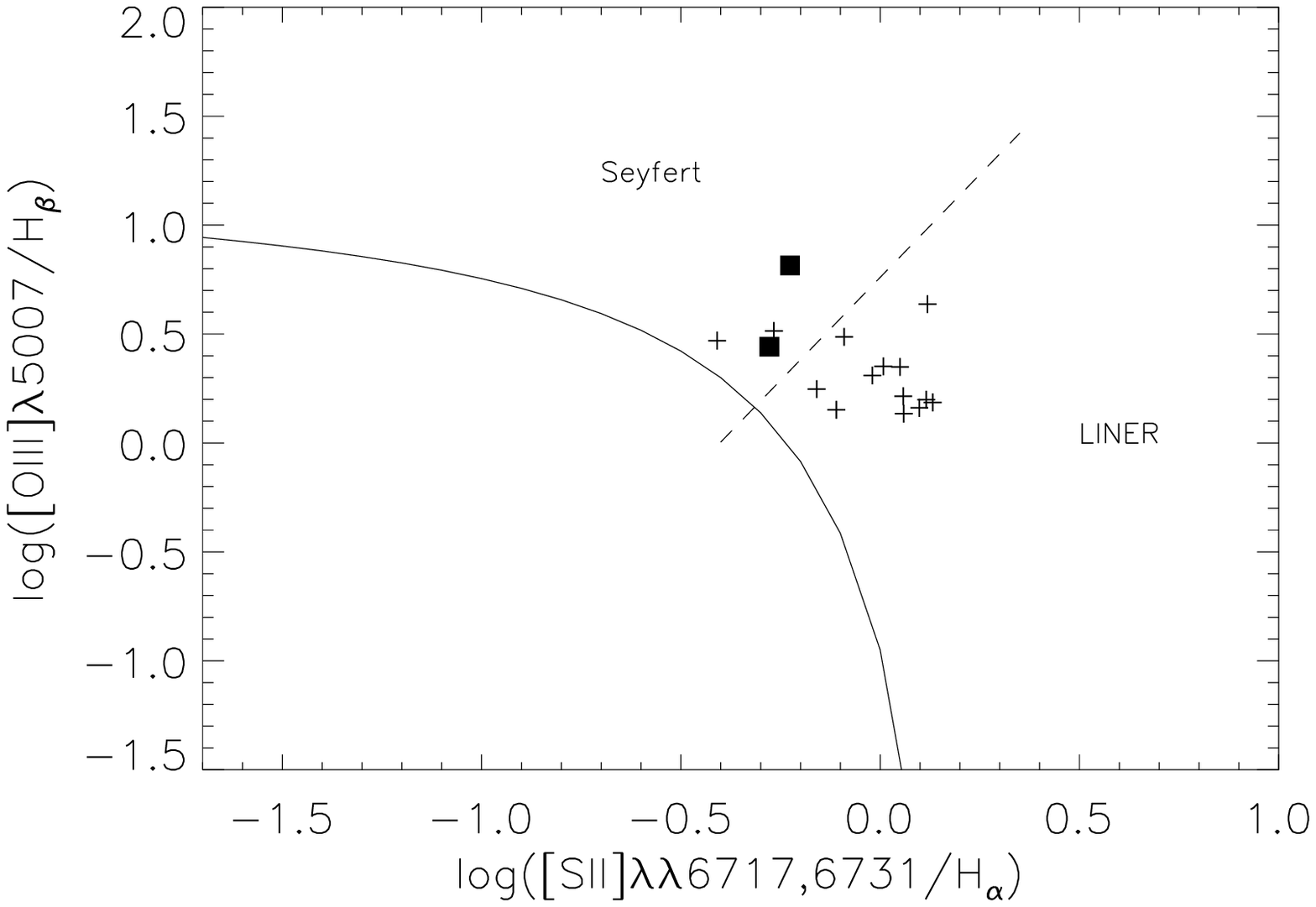}
\includegraphics[width=0.8\hsize,height=0.4\hsize]{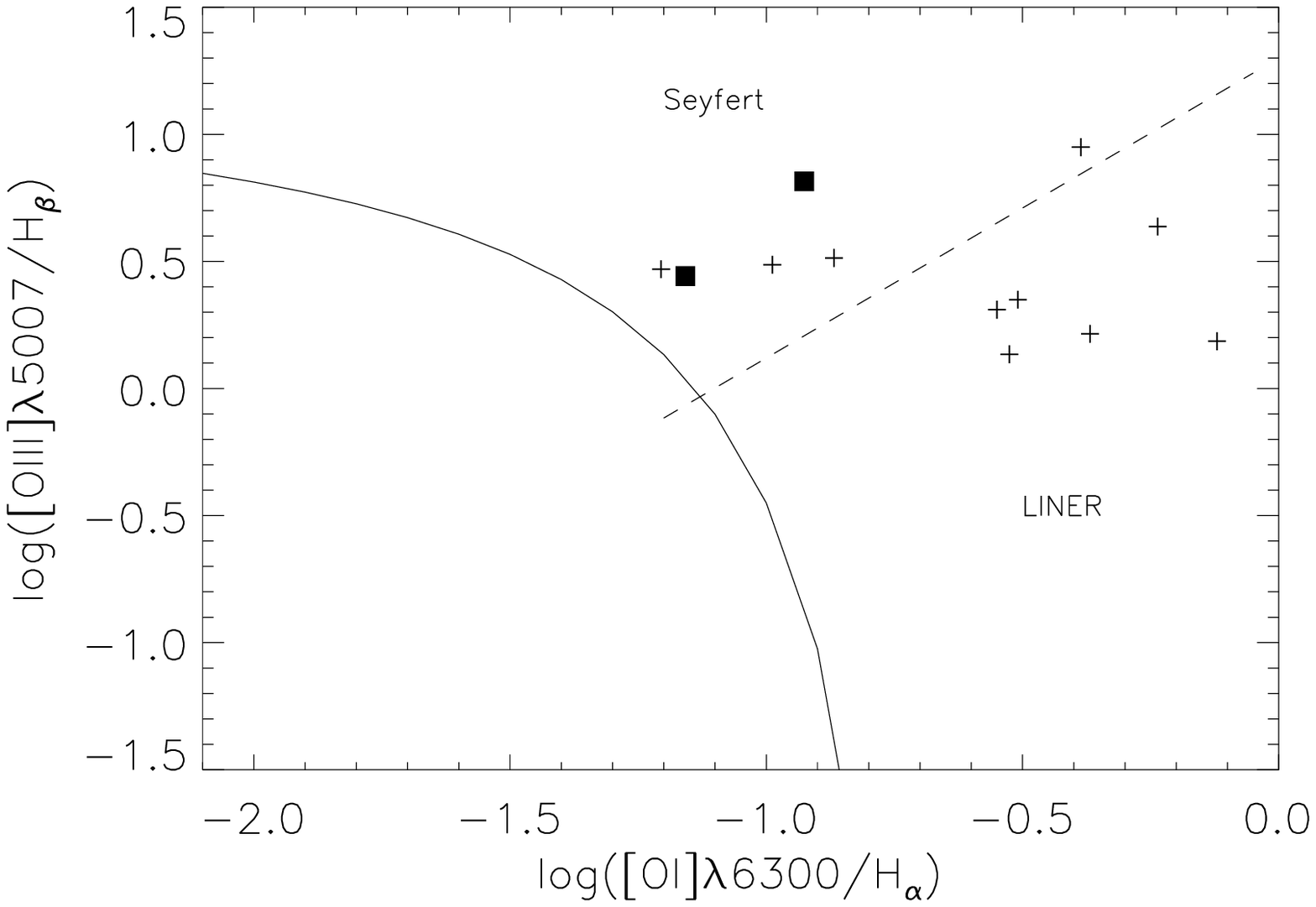}
\caption{The emission line diagnostic diagram based on the AGN classification scheme of Ke06
for 131 emission-line LSBGs with the H$\beta$, [OIII] $\lambda$5007, 
H$\alpha$ and [NII] $\lambda$6583 four lines detected with S/N $>$ 3.
The dashed curve in top panel shows the demarcation between starburst galaxies and AGNs defined by Ke01.
While the solid curve shows the revised demarcation given in Ka03. 
The filled circle in the top panel represents composite object. 
The solid curves in middle and bottom panels represent the 
demarcation between starburst galaxies and AGNs defined by Ke01, 
and the dashed lines represent the demarcation between Seyfert galaxies and LINERs. 
Broad line AGN is marked as filled squares.}
\label{bpt_lsbg}
\end{figure}

The 131 LSB emission-line galaxies are plotted in the 
[NII] $\lambda$6583/H$\alpha$--[OIII]$\lambda$5007/H$\beta$ diagram
in Figure\,\ref{bpt_lsbg} (upper panel), which are marked by
different symbols for different categories.
It can be seen that 
the objects reproduce well the 'butterfly'--shaped distribution as
shown in Ka03, though with much less objects.
The star-forming galaxies (crosses) trace well
the Ka03 dividing line, spreading over a large range of metallicity.
Further separation between Seyferts and LINERs is demonstrated
in Figure\,\ref{bpt_lsbg}, in which the line ratios  
[SII]$\lambda\lambda$6717,6731/H$\alpha$--[OIII] $\lambda$5007/H$\beta$ 
(middle panel)
and
[OI]$\lambda$6300/H$\alpha$--[OIII]$\lambda$5007/H$\beta$ (lower panel)
are plotted for those with the lines detected with S/N$>3$.
For the objects with both the  [SII] and [OI] lines detected with S/N$>3$,
these two criteria give mutually consistent classification. 
Among the three broad line AGN (marked as a filled squares) found in the APM-SDSS sample, 
two have the narrow line properties as Seyferts and
one as composite objects.
The resulting classification of each emission line LSBG 
is given in Table\,\ref{131_lsbgs}.
We also overplot in Figure\,\ref{lsbg_bt_mu0} the distributions of the 
total $B$ magnitudes and \mube\ for the emission line LSBGs.
As demonstrations, in Figure\,\ref{fit_narrow_lsbg} 
we show example spectra and results of spectral
analysis for each of the 3 spectral types. 
We also list in Table\,\ref{impey_30_agn} the host galaxy properties 
of the objects classified as an AGN (Seyfert + LINER + composite),
as given in Impey et al.\ (1996).

\section{Demographics of nuclear activity in LSBGs}

Of the 194 LSBGs, there are 131 ($68\%\pm3$\%) showing emission lines.
Among them, 101 ($52\%\pm3$\%) objects 
are classified as star-forming galaxies and
30 as AGNs ($15\%\pm2$\%; 6 Seyferts, 12 LINERs, and 12 composite galaxies).
The fraction of broad line AGNs is $2\%\pm 1\%$ among LSBGs.
This result indicates that the fraction of AGNs is significantly lower
than that ($>$40$\%$) found for the local bright galaxy sample of Ho et al.\ (1997a),
which is  predominantly HSBG. 
The fraction of AGNs is even lower if not all LINERs and composite objects are AGNs.

As discussed in the sample selection above, probably not all the 
194 LSBGs are low surface brightness in a strict sense.
In fact, 79 (40$\%$) of the 194 LSBGs have \mube\ 
from \citet{b12} brighter than 22.0 mag arcsec$^{-2}$,
which are either LSBGs with (dominating) galactic bulges or simply HSBGs. 
Among them, there are 25 AGNs, yielding an AGN fraction of $31\%$.
Considering possible contamination of HSBGs
in this bright subsample, this fraction is likely overestimated
and the actual fraction for genuine LSBGs is likely even lower.
The remaining 115 galaxies with \mube$>22.0$\,\umu should be mostly
bona fide LSBGs (Impey01).
Among them, there are only 5 AGNs, making a fraction as low as 4$\%\pm2\%$.
This strengthens the above result that LSBGs have a much lower AGN
fraction compared to HSBGs.

\begin{figure}
\includegraphics[width=0.5\hsize,height=0.6\hsize]{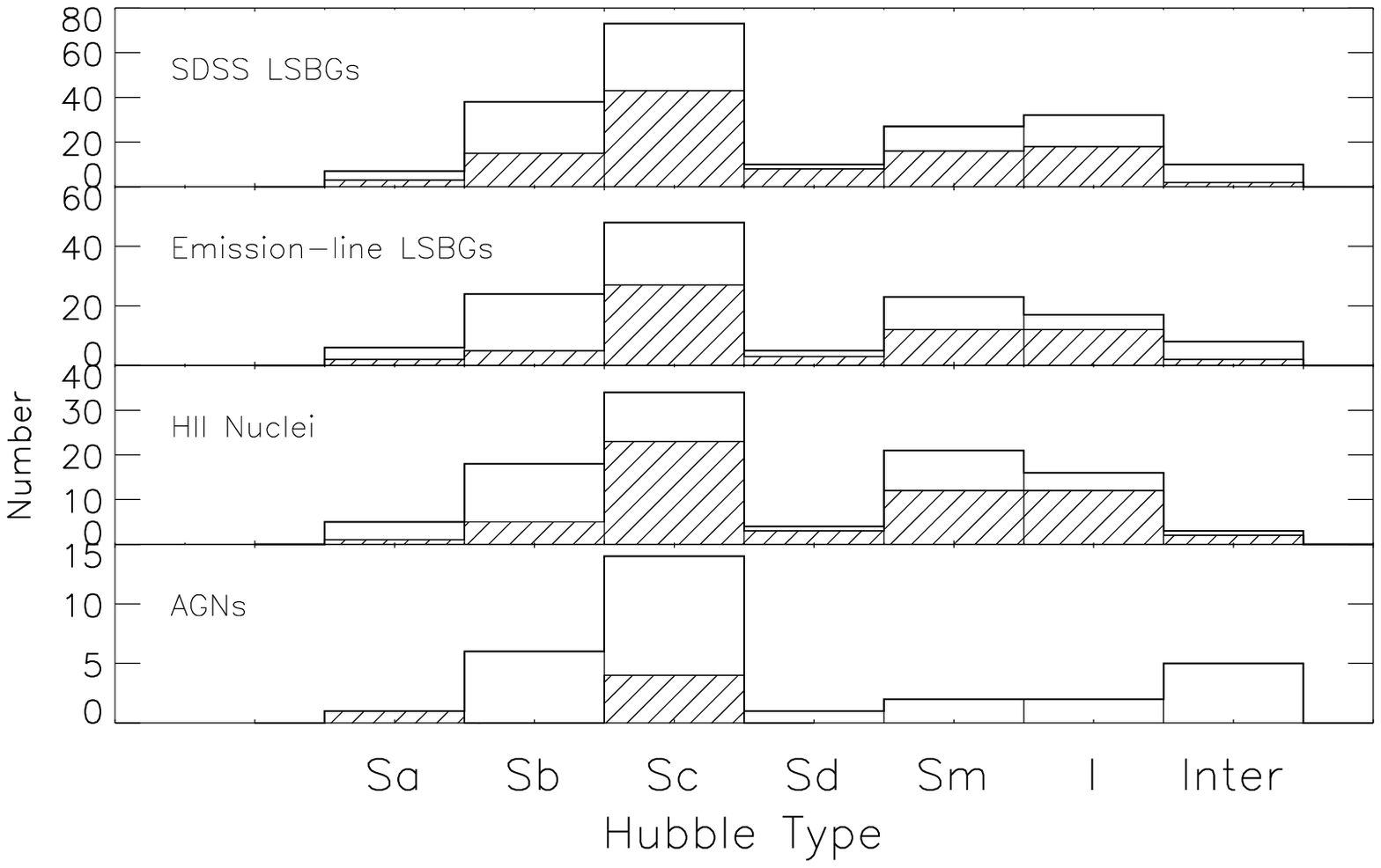}%
 \includegraphics[width=0.5\hsize,height=0.6\hsize]{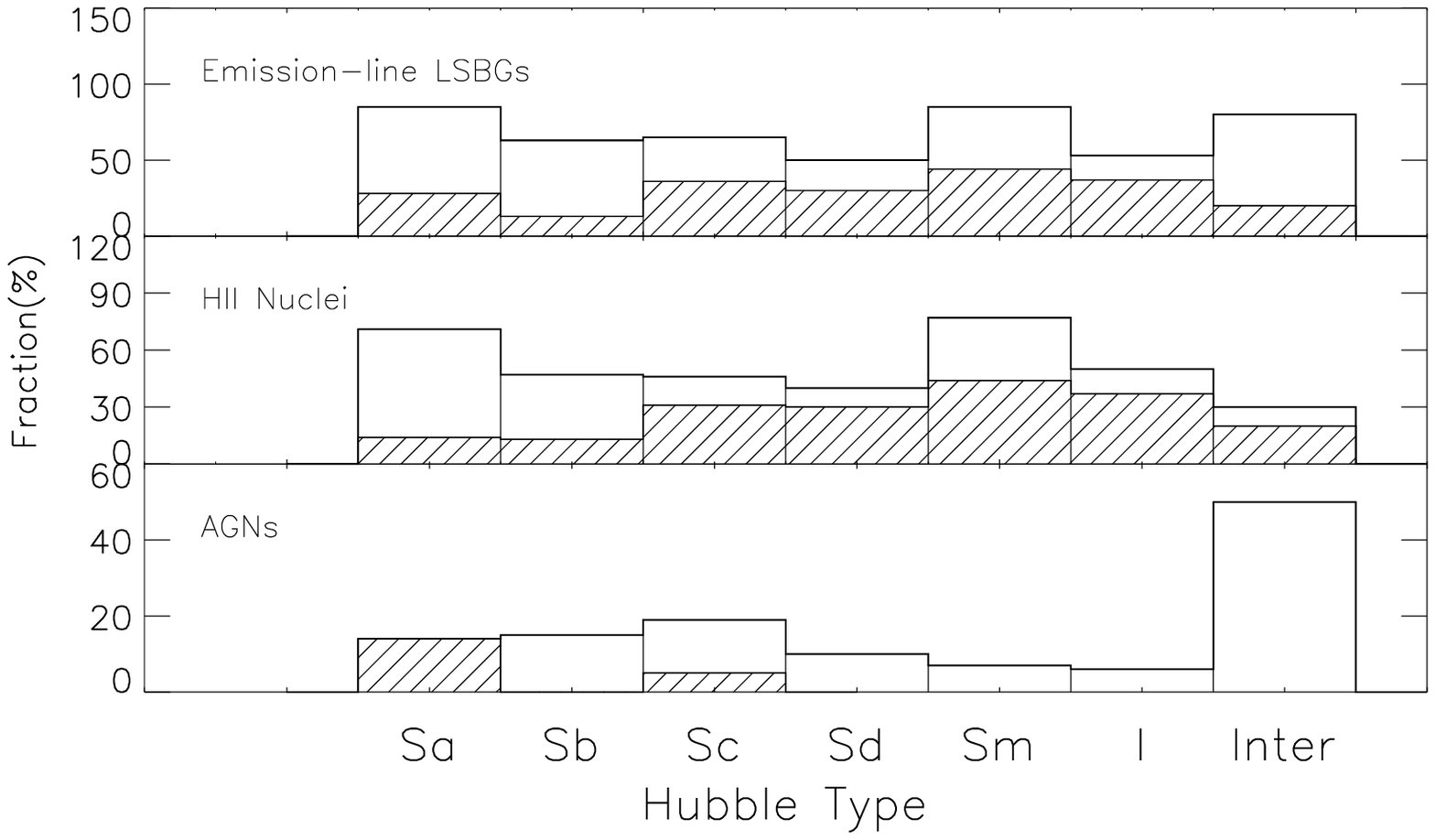}
\includegraphics[width=0.5\hsize,height=0.6\hsize]{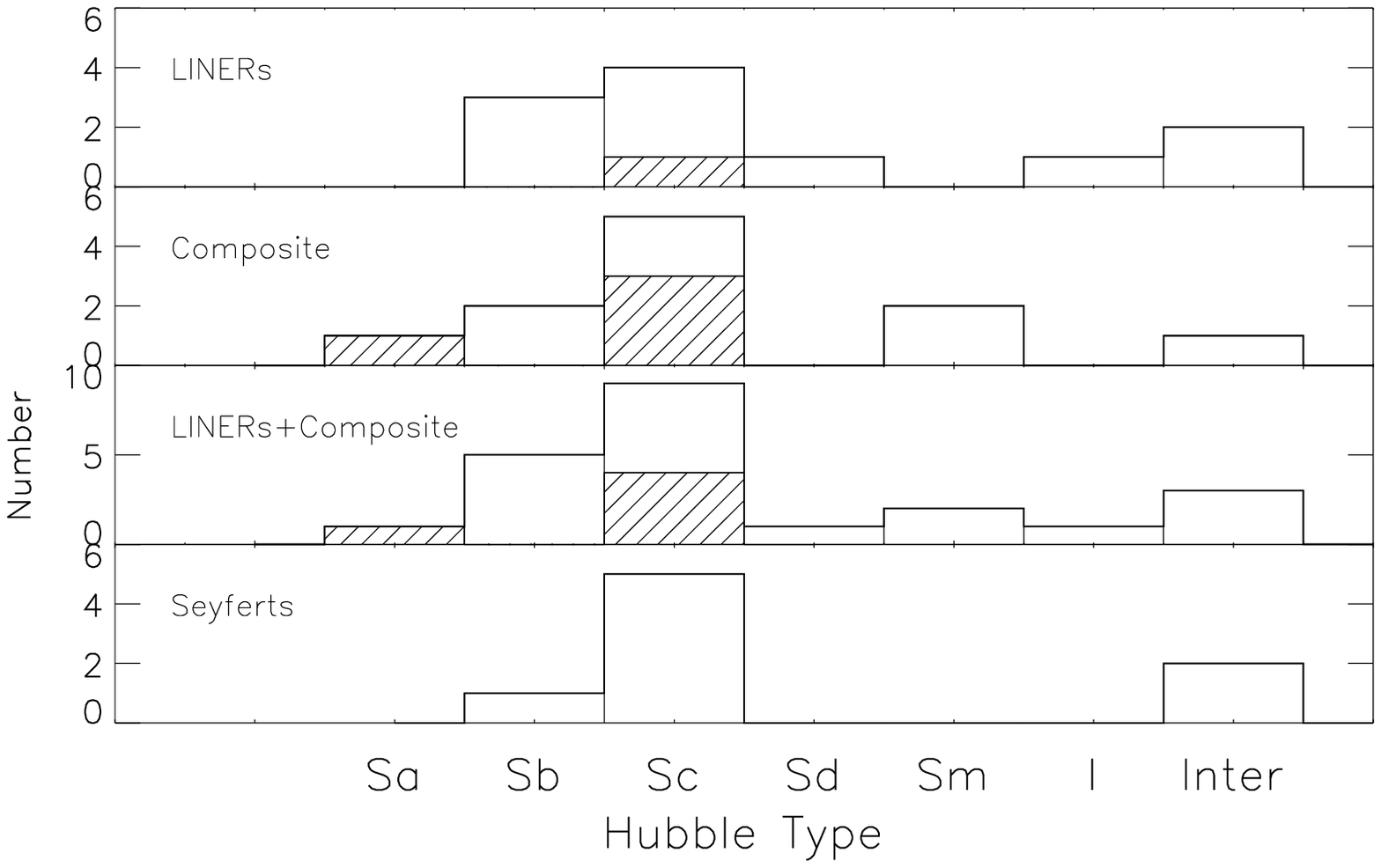}%
 \includegraphics[width=0.5\hsize,height=0.6\hsize]{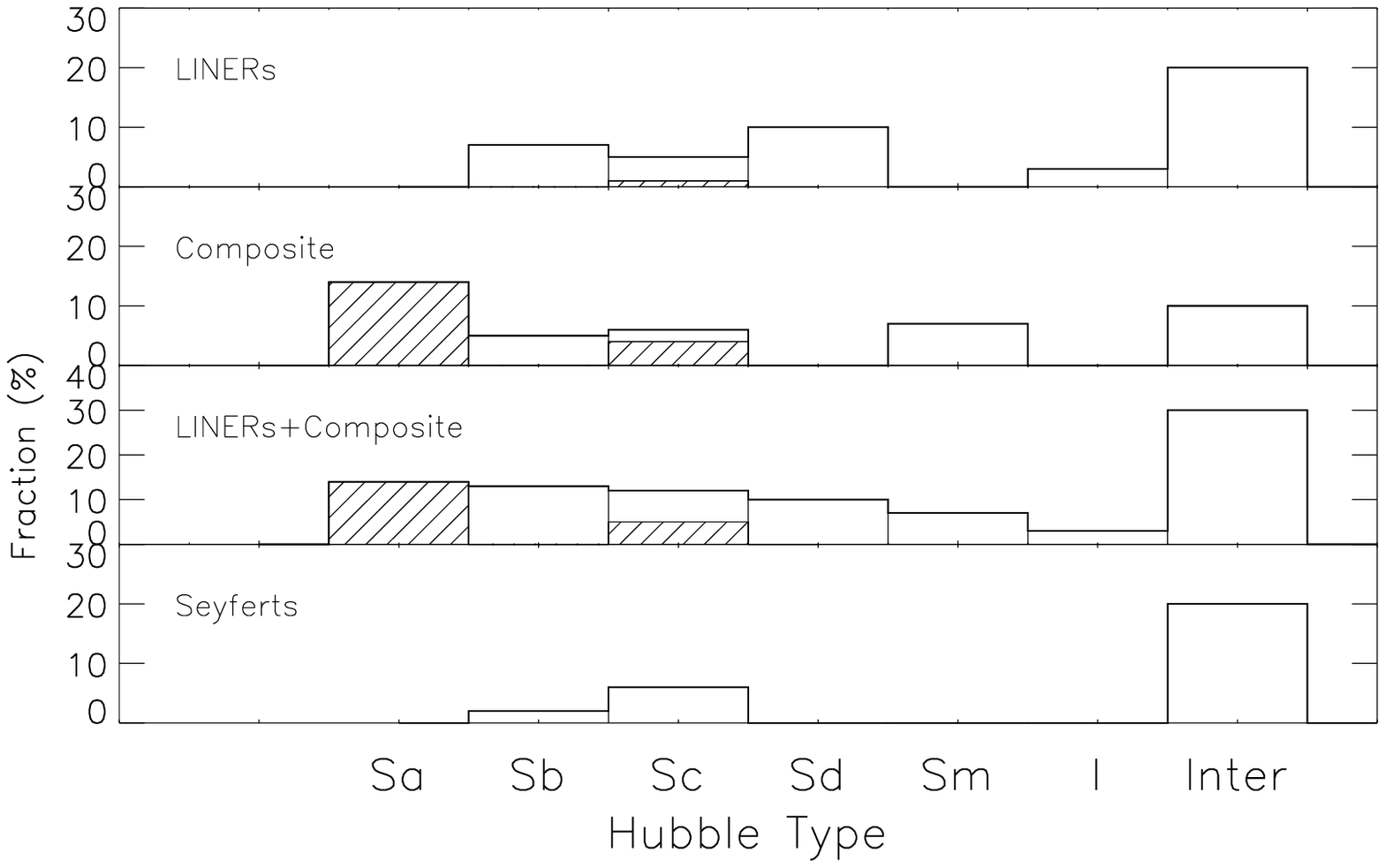}
\caption{Number statistics and detection rates of LSBGs as a 
function of morphological types of all emission-line nuclei,
 HII nuclei, and AGNs. LSBGs with 
\mube $\geq$ 22.0 mag arcsec$^{-2}$ are shown by hatched histogram.}
\label{spiral_lsbg}
\end{figure}

The incidence of an AGN has been found to be dependent on the Hubble type
of a galaxy, that is higher in early type (E--Sb) than
in late type galaxies (Ho, et al.\ 1997a).
Since a LSBG sample has a very different Hubble type distribution from HSBGs
(the former tends to consist of more late type than the latter),
the above result may be a reflection of the incidence of AGNs on galaxy morphology.
To investigate this possibility, we calculate the fraction of both AGNs and
star-forming galaxies in each morphological type,
using the Hubble type information provided in \citet{b12}.
The distribution of the objects of each category and their detection rates
over the morphological type are shown in Figure.\ref{spiral_lsbg} (open histograms).
Also plotted are the same distributions for the \mube$>22.0$\,\umu subsample
(hatched histograms).
The numbers are also given in Table\,\ref{frac_mu}.
It can be seen that the LSBGs have most morphological types
later than Sb and peak at Sc, as expected.
The detection rate of AGNs is no more than 20\% in LSBs of
type Sa, Sb, and Sc, and
starts to drop to less than 10\% in types later than Sc.
This result indicates that the overall low AGN fraction of LSBGs
than that of HSBGs is real, rather than a consequence that LSBGs
tend to have more of later Hubble types, which have in general
low AGN fractions, as known previously.
Interestingly, it  increases up to $50\%$ in interacting galaxies.
When galaxies with  \mube$>22.0$\,\umu
are considered only, the detection rates are even lower,
though the sample is small.

\subsection{Comparison with HSBGs}

\begin{figure}
\includegraphics[]{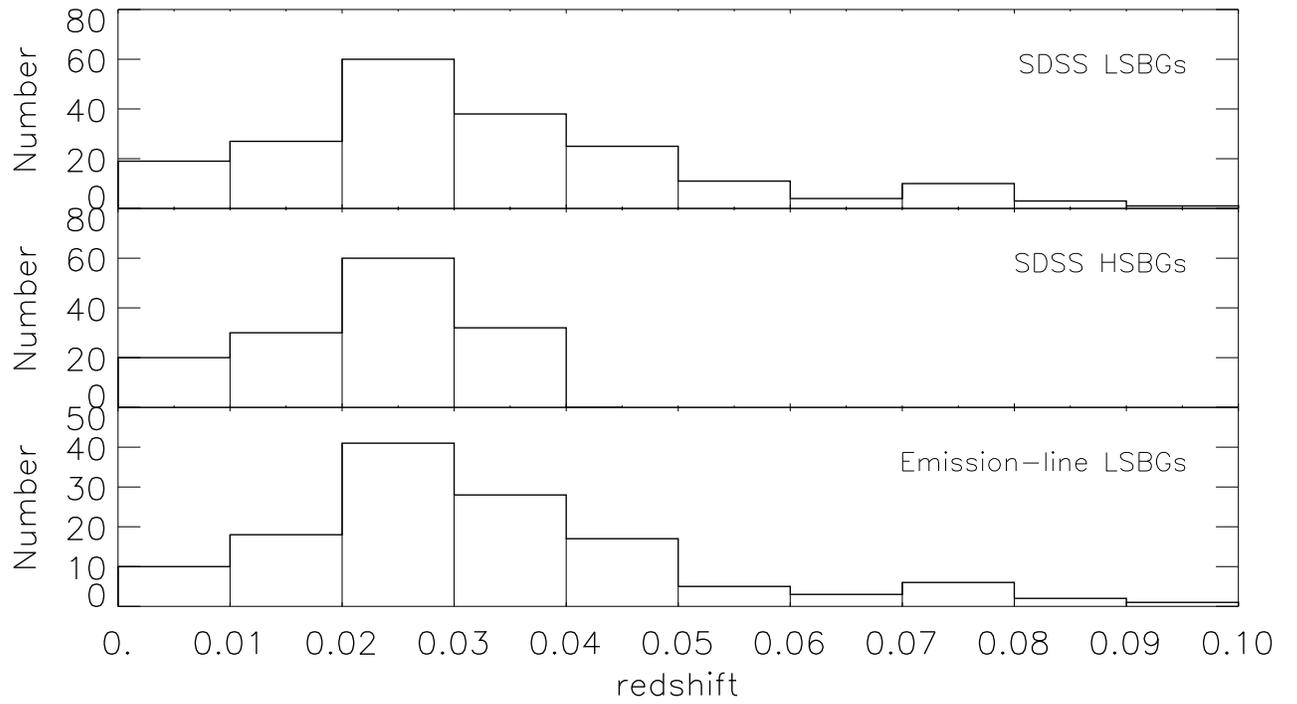}
\caption{Distribution of redshift for the APM-SDSS LSBGs (top panel), 
the comparing HSBG sample (middle panel) and the 131 emission-line LSBGs (bottom panel).}
\label{z_dis}
\end{figure}

Strictly speaking, an appropriate comparison of AGN detection rate between
two samples requires that they should have the same redshift
(distance) distributions, and have the same spectral data quality
and analysis procedures.
To make sure that our above results are not significantly affected
by those biases introduced in sample selection and observation,
we construct a comparing sample of HSBGs with SDSS spectra.
The sample is drawn from
the Third Reference Catalogue of Bright Galaxies (RC3) \citep{b64}---a
catalog of typical HSBGs--in such a way that it has the same
redshift (distance) distribution as that of the APM-SDSS LSBG sample
below $z=0.04$\footnote{This redshift cutoff is introduced because
there are few RC3 galaxies with $z>0.04$}.
This results in a comparing sample consisting of 142 HSBGs
with SDSS spectra.
For a comparison, the redshift distribution of the HSBG sample is plotted in
Figure\,\ref{z_dis} (middle panel), which is indistinguishable from that of the LSBGs
in the $z=0-0.04$ range (the K-S test gives a chance probability
of 0.1 that the two distributions are drawn from the same population).
The HSBG sample is systematically brighter than the LSBG one
in total magnitude, as expected.
The SDSS spectra of the 142 HSBGs are analyzed using the exactly
same procedure as for the LSBGs in Section\,2., ensuring the
homogeneity  in data analysis.

\begin{figure}
\includegraphics[]{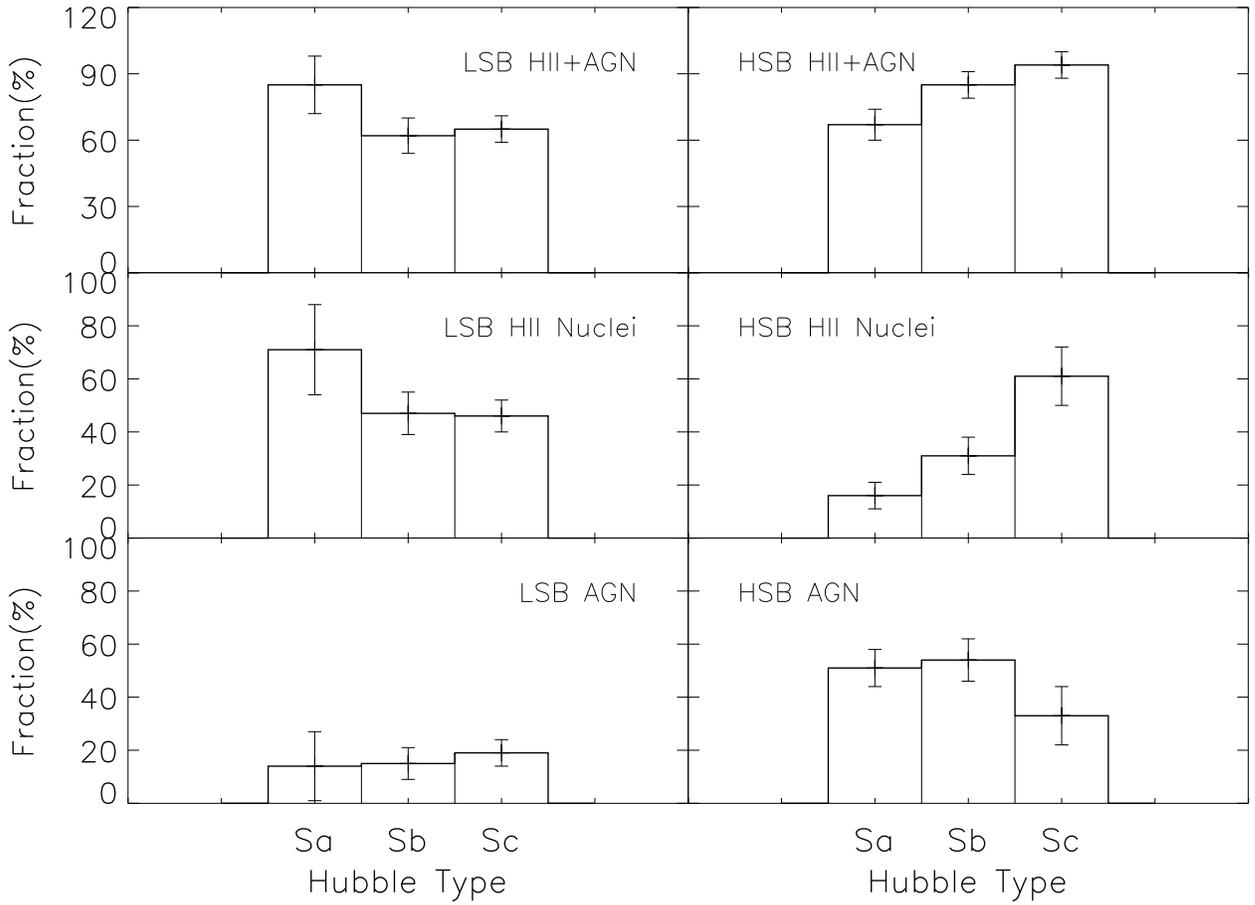}
\caption{Fraction of galaxies containing star-forming and AGN nuclei for the 
LSBGs with z $<$ 0.04 (left panels)
and the comparing HSBG sample (right panels),
for the Sa, Sb and Sc types.}
\label{lsb_hsb_frac}
\end{figure}

We compare the  LSBG and HSBG samples in the range of $z<0.04$,
and only those with the morphological types of  Sa, Sb, and Sc are considered
in order to have sufficient objects to ensure statistically meaningful results.
The detection rates of AGNs and star-forming galaxies for both the
LSBGs and HSBGs are shown in Figure\,\ref{lsb_hsb_frac}.
The results for  HSBGs (right-hand side panels)
are broadly consistent with those obtained by  Ho et al.\ (1997a),
suggesting that our results are reliable.
Comparisons between the LSBGs (left-hand side panels) and HSBGs
show clearly  that LSBGs indeed have a low AGN fraction
(10--20\%) than HSBGs (40--50\%), regardless their morphological types.
The numbers are also given in Table\,\ref{lsb_hsb_Sx}.
Thus our above results are confirmed.

Interestingly, the fraction of star-forming galaxies in the LSBGs seems to
be comparable to, or even slightly higher than, that in the HSBGs.
A surprising difference is found in the distribution of the fraction
of star-forming galaxies: in HSBGs, this fraction drops significantly
from Sc to earlier types and is the lowest for Sa.
While in LSBGs, it keeps a high fraction toward early types
and even possible rises for the Sa type.

Summarizing, our results do not apparently support the high AGN fraction
reported in Sprayberry et al. (1995) and Schombert (1998),
but are consistent qualitatively with that of Impey01.

\section{Properties of AGNs in LSBGs}

\subsection{Power of AGN}

\begin{figure}
\includegraphics[scale=0.7]{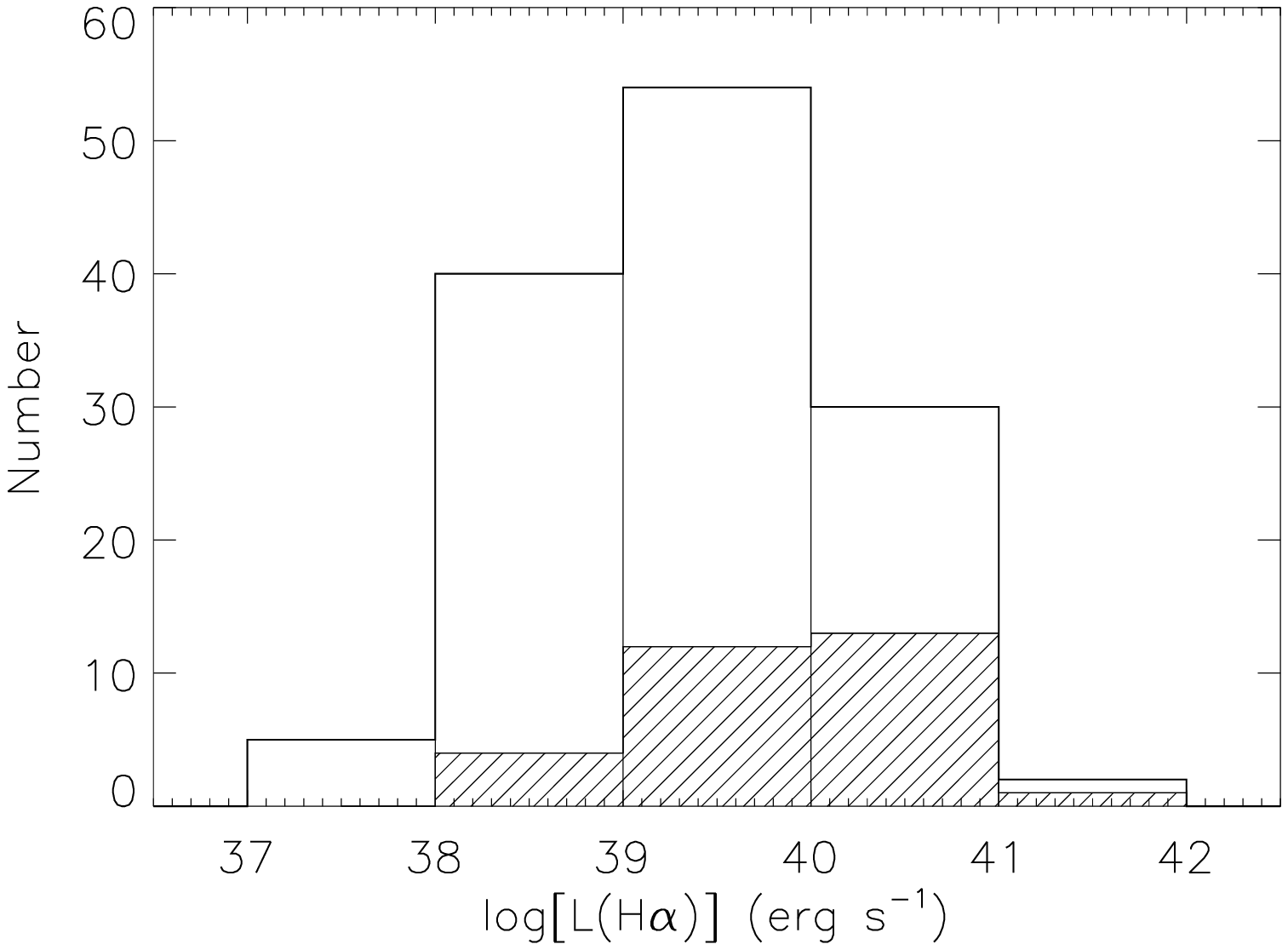}
\includegraphics[scale=0.7]{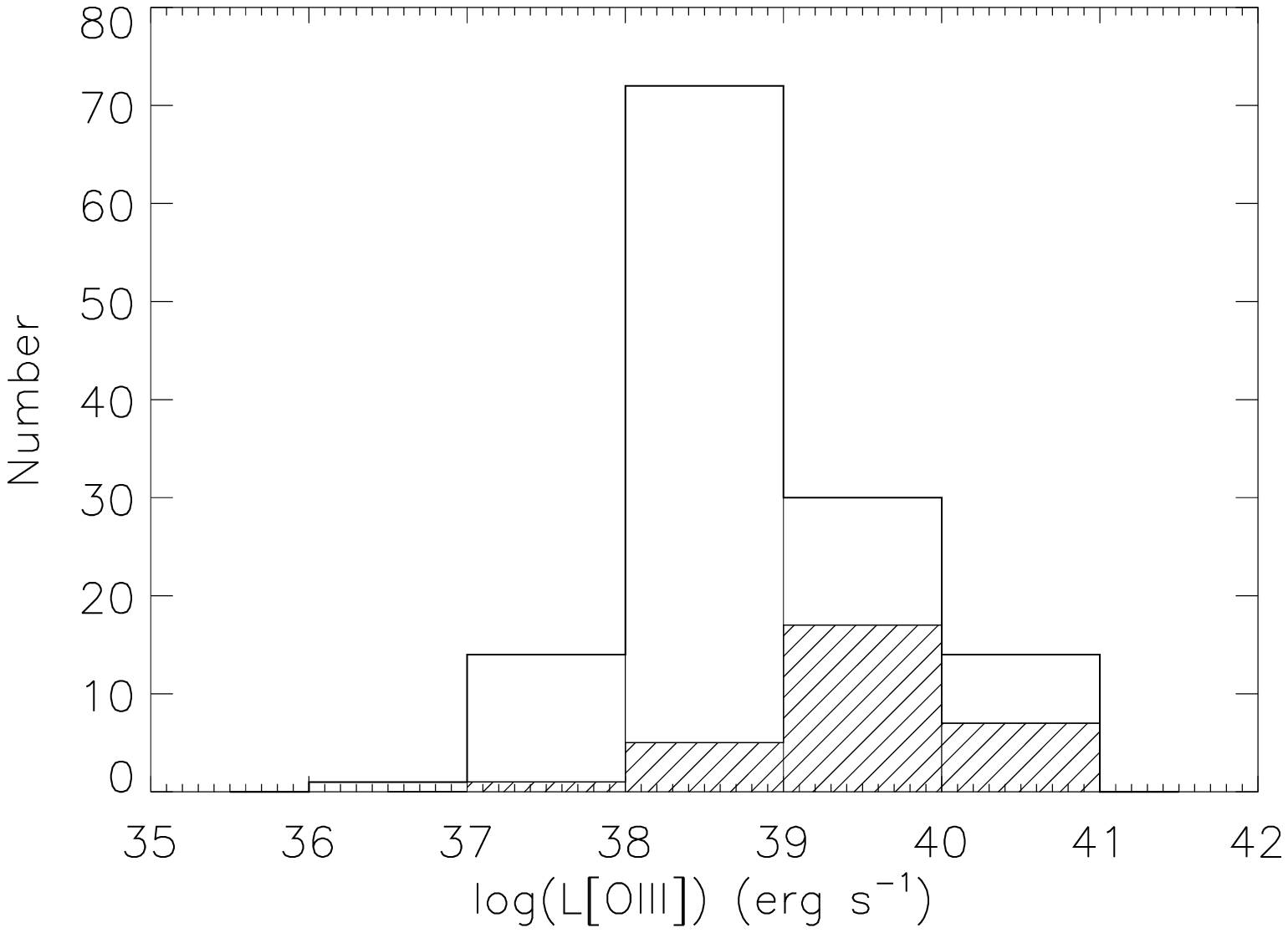}
\caption{Distributions of luminosities of the narrow H$\alpha$ line 
(in units of erg s$^{-1}$) and [OIII] $\lambda$5007 line of LSBGs. 
That the higher [OIII] $\lambda$5007 luminosities, 
the higher fractional AGNs indicates that [OIII] $\lambda$5007 
luminosities are correlated with the strength of AGN activities. 
The hatched histogram denotes AGNs.}
\label{ha_oiii_lum}
 \end{figure}

Figure\,\ref{ha_oiii_lum} shows the distributions of
the H$\alpha$ and [OIII] emission line luminosities of the APM-SDSS LSBG sample.
The majority of the AGNs
have H$\alpha$ luminosities lower than 10$^{40}$ ergs s$^{-1}$,
with a median of 7.5 $\times$ 10$^{39}$ ergs s$^{-1}$.
All of the AGNs in  LSBGs have
L[OIII] $<10^{41}$ ergs s$^{-1}$
and the distribution peaks between 10$^{38}$--10$^{39}$ ergs s$^{-1}$. 
These values are 2 to 3 orders of magnitude smaller than those
of luminous AGN.
It has been suggested that the luminosity of the [OIII]$\lambda$5007
line can be regarded as a tracer of AGN activity (Ka03).
Therefore, the AGNs in LSBGs are mainly of low luminosity.
Interestingly, as inferred from Figure.\ref{ha_oiii_lum},
the AGN detection rate becomes higher with higher L[OIII], i.e.
increasing gradually from $\sim$7\%
at 10$^{37}$ ergs s$^{-1}$  up to $\sim$78\% at 10$^{41}$ ergs s$^{-1}$.

\subsection{Properties of emission line region}

\subsubsection{Narrow line width and stellar velocity dispersion}

\begin{figure}
\includegraphics[]{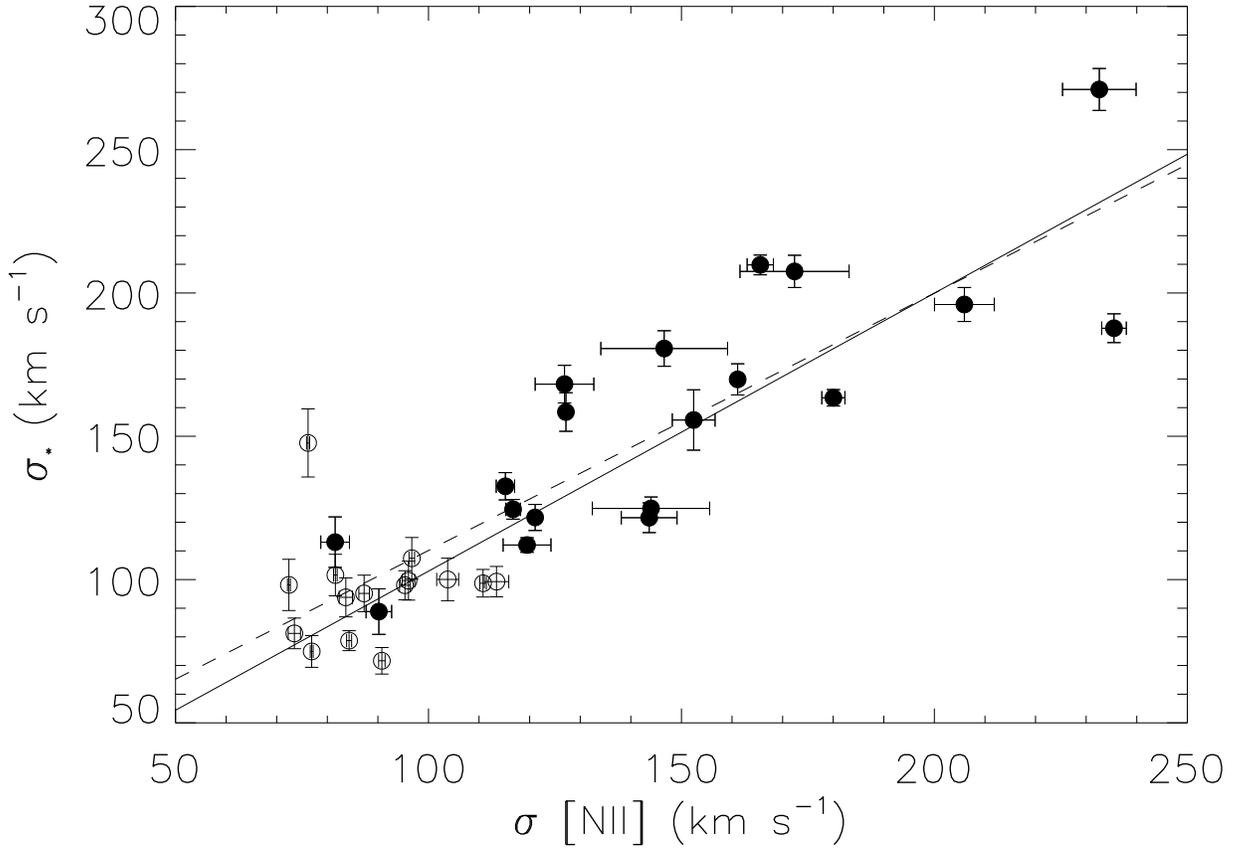}
\caption{Relationship between the width of the [NII] emission line 
and the galactic stellar velocity dispersion $\sigma_*$ for 
34 LSBGs which have both measurements of $\sigma$[NII] and 
$\sigma_*$ with S/N $>$ 5, including 19 AGNs (filled-circles) and 15
star-forming galaxies (open circles). Also plotted are linear relations 
for all the 34 LSBGs (solid line) and for AGNs only (dashed line) 
fitted by the regression analysis taking into account the errors 
in both the variables.}
\label{nii_sigma_star}
\end{figure}

In AGNs of normal HSBGs, it has been suggested that kinematics of
the narrow line region traces well the galactic bulge potential,
via the establishment of a tight correlation between the
width of the narrow emission lines and the stellar velocity dispersion $\sigma_\ast$.
Here we test this relationship for AGNs in LSBGs.
The value of $\sigma_\ast$ can be obtained in our modeling of the
stellar spectra of the host galaxy starlight (Section\,2).
For 34 LSBGs (19 AGNs and 15 star-forming galaxies) with both $\sigma_\ast$ and the [NII] doublet
detected at the $>5\,\sigma$ level,
the relation between $\sigma_\ast$ and the [NII] line width
is plotted in Figure\,\ref{nii_sigma_star}.
Strong correlations are found 
(For AGNs, $\sigma_\ast$ $=$ 0.9 $\times$ ($\sigma$[NII] $-$ 150.0) $+$ 155.3 
; for the whole objects, 
$\sigma_\ast$ $=$ 0.97 $\times$ ($\sigma$[NII] $-$ 150.0) $+$ 151.4. The Spearman correlation test gives
a chance probability of 8.6 $\times$ 10$^{-6}$ for AGNs only and 1.5 $\times$ 10$^{-9}$ for AGNs and star-forming galaxies.)
Thus, the width of the narrow line $\sigma$[NII]
traces well the stellar velocity dispersion in LSBGs.

\subsubsection{Electron density in the narrow line region}

\begin{figure}
\includegraphics[]{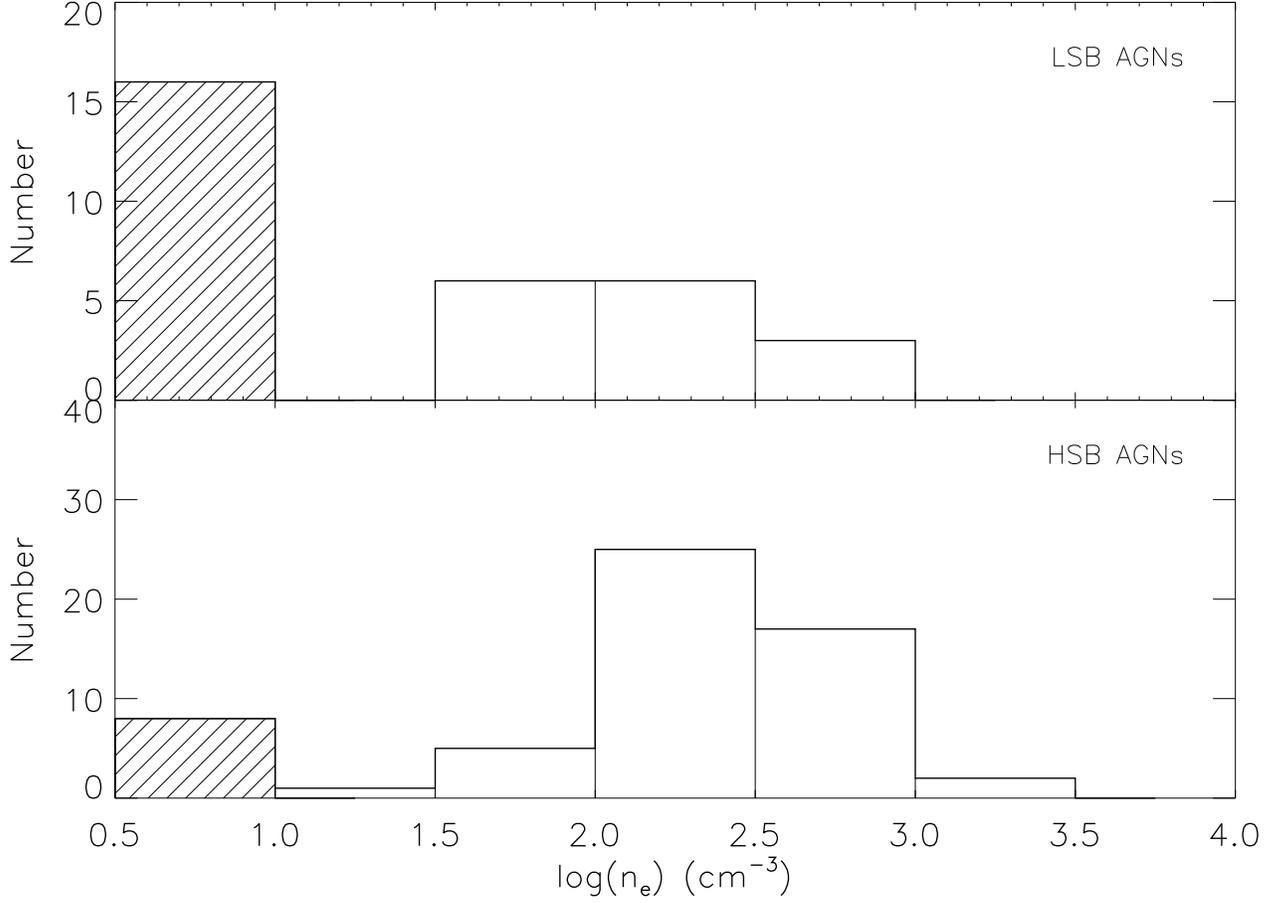}
\caption{Distribution of the narrow$-$line region (NLR) 
electron densities of AGNs in LSBGs (upper panel) and in the comparing sample of HSBGs (lower panel). 
An upper limit of n$_{e}$ = 10 cm$^{-3}$ is set to those with 
F$_{[SII]\lambda6717}$ / F$_{[SII]\lambda6731}$ greater than 1.42, which are shown as hatched histogram. 
AGNs in LSBGs tend to have lower NLR electron densities than those of HSBGs 
(the probability that the two samples have the same distribution is 
P$_{ks}$ = 7.0 $\times$ 10$^{-5}$).}
\label{e_density}
\end{figure}

Since in the course of our spectral analysis the
[SII]$\lambda\lambda$6717,6731 doublet is deblended,
we can assess the electron density in the narrow line region (NLR)
from the flux ratio of these two lines (e.g., Osterbrock \& Ferland 2006; Xu et al 2007).
We assume an electron temperature of T$_{e}$ = 10$^{4}$\,K,
a typical temperature of the ionized gas in the NLR of AGNs.
The electron densities of AGN in both  the LSBG and HSBG sample
are estimated, whose distributions are shown in Figure.\ref{e_density}.
For those with [SII]$\lambda$6717/[SII] $\lambda$6731 flux ratios
greater than 1.42, an upper limit of the electron densities is set to be
10\,cm$^{-3}$.
A comparison shows that these two distributions are significantly different
(a chance probability of $7\times 10^{-5}$ given by the K-S test),
in a way that AGNs in LSBGs on have average lower NLR electron density
than AGNs in HSBGs.

\subsubsection{Dust extinction}

We examine optical extinction in the emission line nuclei of the LSBGs
by assuming that the Balmer decrements in excess of the
intrinsic value 2.85
for HII region and 3.1 for AGNs \citep{b41}, are caused by dust extinction.
The narrow lines Balmer decrements are calculated for star-forming galaxies
and AGNs for both the LSBGs and the HSBGs.
For the LSBGs, the median Balmer decrement is 3.3 for star-forming nuclei
and 4.0 for AGNs, which correspond to
an extinction color excess E$_{B-V}$ = 0.16 and 0.24, respectively.
For the HSBG comparing sample,
the median values are 4.2 for  star-forming nuclei and 4.7 for AGNs,
which correspond to E$_{B-V}$ = 0.37 and 0.40, respectively.
This indicates that dust extinction in LSBGs is relatively
weaker compared to that in normal galaxies, for both AGNs and star-forming nuclei.
A comparison of the distributions of the Balmer decrement between
the HSBGs and the LSBGs shows that they are statistically different,
for both AGNs and star-forming nuclei 
(a chance probability of $2.7\times 10^{-7}$ given by the K-S test for AGNs and star-forming galaxies).

\subsection{Radio and X-ray detectability of AGNs in LSBGs}
   Observationally, AGNs emit their power over a wide range of frequencies,
from radio, optical, to X$-$rays or even $\gamma$$-$rays.
Among the 30 LSBG AGNs, only one, J231815.7$+$001540.2 (NGC\,7589)
was detected in X-rays with XMM-Newton,
which exhibits large amplitude variability \citep{b24}.
For the other AGNs, the X-ray flux limits set by the ROSAT All-Sky Survey
are comfortably consistent with the distribution of the
optical-to-X-ray effective spectral indices $\alpha_{ox}$ found for AGNs.
Deeper X-ray observations are needed to detect more such AGNs in X-rays.

Of the 30 AGNs in LSBGs, 8 (26$\%$ $\pm$ 8$\%$) were detected by the FIRST
(Faint Images of the Radio Sky at 20 cm) survey \citep{b50}),
including one with broad emission lines.
Among them, 2 are resolved in their radio images.
We calculate the radio loudness R,
defined as the flux ratio between 1.4 GHz and the optical B-band.
Six objects have $R>10$ and can be classified as radio-loud.
Most of the radio-loud objects of our LSBG AGNs are actually radio-intermediate
($R<100$), and there is no very radio loud (R $\geq$ 100) AGNs.
They are all weak radio sources, with 1.4GHz fluxes  less than 10 mJy.
For the remaining 22 AGNs, 20 were in the FIRST survey field.
Assuming an upper flux limit of 1\,mJy for the FIRST survey,
we found that at least 13 are "radio-quiet" (R $\leq$ 10).
This makes the fraction of radio-loud fraction as 20\%--35\%,

\section{Discussion}

\subsection{AGN fraction in LSBGs}

We have shown that LSBGs have a lower AGN detection rate
compared to HSBGs in the local Universe.
The overall  fraction of AGNs in the APM-SDSS LSBG
sample is $\sim15\%$. Considering that the APM LSBG sample have
possible contamination of some HSBGs,
the AGN fraction in genuine LSBGs can be even lower.
This result holds true within each of the Hubble types from Sa to Sc.
We argue that this result is not caused by any selection effects
or detecting biases in observations, 
since it is confirmed by comparisons with
the carefully chosen HSBG comparing sample.
In fact, AGNs in LSBGs should be more easy to detect spectroscopically
compared to those in HSBGs, since the dilution of AGN spectral features 
by host galaxy starlight spectra is less severe in LSBGs than in HSBGs.
To check whether we have missed a substantial amount
of AGNs at larger distances, we examine how the AGN detection rate
varies with increasing distances. We find that the AGN detection rate
keeps nearly a constant at all distances within the largest redshift of th sample ($z<0.1$), 
and thus there is no significant number of faint AGNs missing at large distances.

In general, a low AGN occurrence rate 
may be possibly due to the inefficiency of 
fueling gas into the central black holes in galaxies.
However, the underlying physical processes responsible for
the fueling efficiency is unknown, which is one of the fundamental
questions in AGN research.
One speculation is that  gas in LSBGs may have relative 
large  angular momentum \citep{b58},  even in the inner disk.
Alternatively, it may well be possible that this result is just a manifestation 
of the more fundamental differences  between  HSBGs and LSBGs 
in some of the physical properties in the central region of
galaxies (such as bulges), which are more directly 
linked to the onset of nuclear activity.

\begin{figure}
\includegraphics[width=0.8\hsize,height=0.7\hsize]{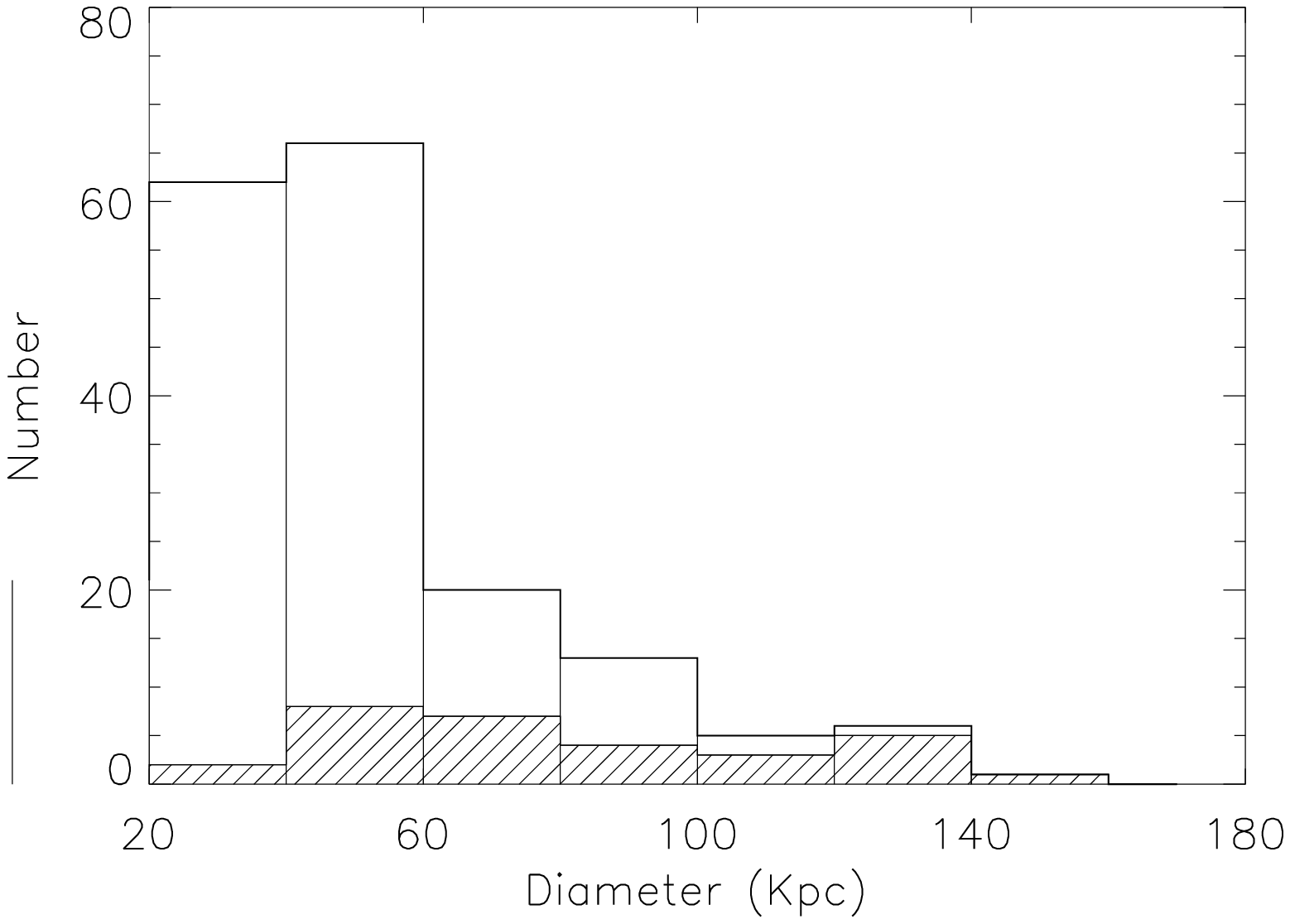}
\includegraphics[width=0.8\hsize,height=0.7\hsize]{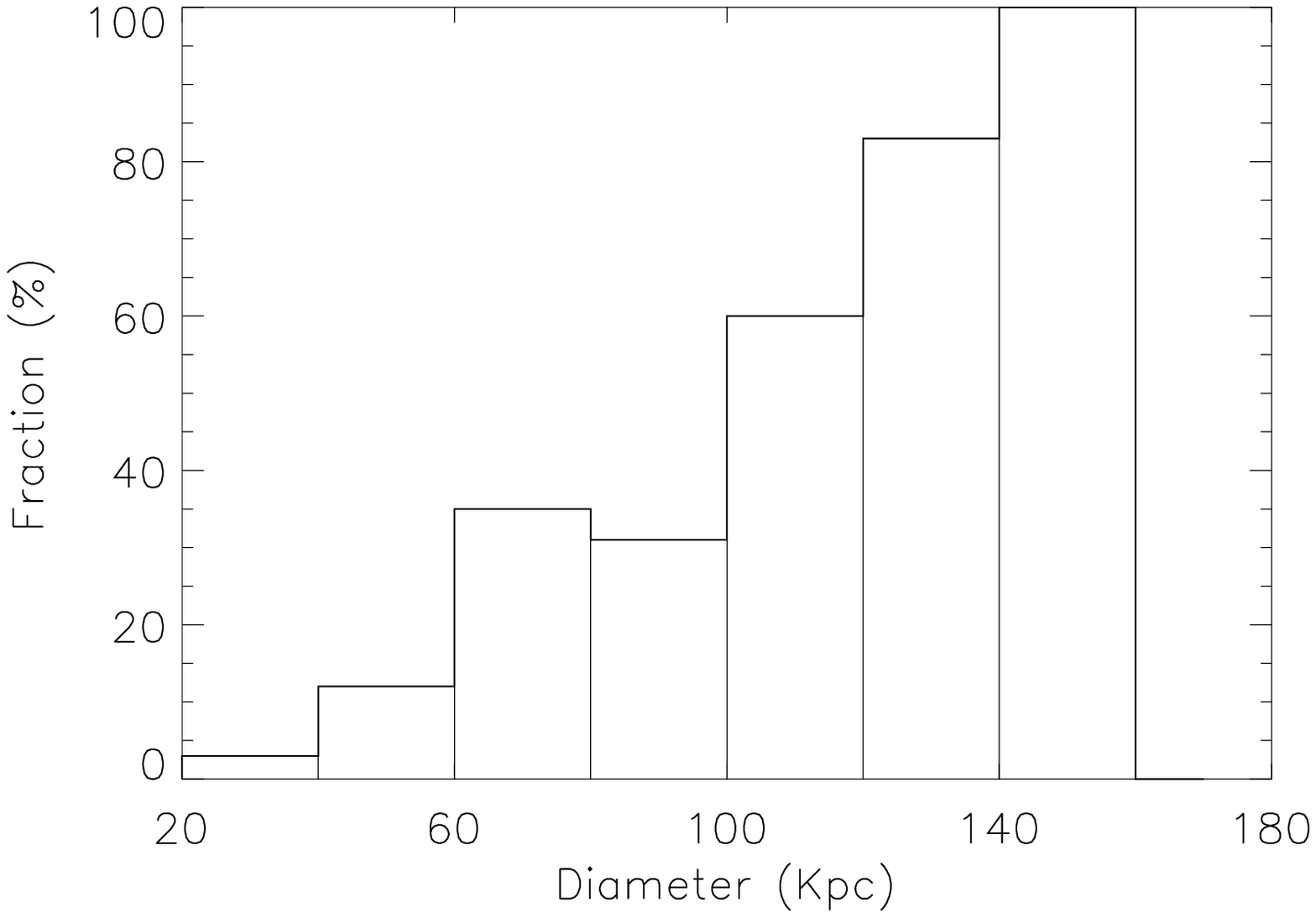}
\caption{Distribution of the physical sizes of LSBGs (upper panel,
AGNs are plotted as hatched histogram) and the fractions of AGNs as a function of the galactic size (lower panel).}
\label{agn_frac_diam}
\end{figure}

Motivated by this hypothesis, we search for other potential
dependence of the AGN occurrence rate in the LSBG sample.
One such dependence is on the physical size of galaxies.
Figure\,\ref{agn_frac_diam} shows the distribution of 
the size of galaxies in the APM-SDSS LSBG sample 
and the dependence of the AGN fraction on the galaxy scale.
The physical scales of the LSBGs are estimated using the 
angular sizes measured in the SDSS imaging survey by the SDSS pipeline.
As can be seen, there exists a clear trend that 
larger galaxies tend to have a higher fraction of AGNs.
This implies that the overall AGN fraction for a sample of LSBGs
depends on the distribution of the physical scales of galaxies
in the sample. 
For the current APM-SDSS LSBG sample, 
the vast majority of the member galaxies are smaller than 50\,kpc,
which have very low fractions of AGNs.
This finding also implies that, when AGN fractions are compared
between two LSBG samples, the distributions of
the physical scales have to be taken into account. 

\citet{b21} reported a high occurrence of AGNs ($\sim 50\%$)
in his LSBG sample, which 
are mostly late-type, HI-massive,  and large-size galaxies.
In fact, using the data presented in \citet{b21},
we also find a similar strong dependence of the AGN fraction on 
galaxy size.
We compare the distributions of galactic sizes between the \citet{b21} sample
and the APM-SDSS sample; however,
no significant difference is found 
(the K-S test gives a probability of 68$\%$).
Thus, we anticipate that the higher AGN fraction found in 
\citet{b21} than ours is perhaps due to the small sample size,
as well as possibly to the relatively coarse spectral quality, and
less rigorous spectral analysis and classification method employed.

A similar trend of increasing AGN fraction with the increase
of the size of a galaxy is also found for HSBGs  in the comparing 
sample. Meanwhile, we also find that the HSBG sample has relatively
systematically larger galaxies in size (with a  median of 42.1\,kpc) 
than the LSBG sample (with a median of 37.4\,kpc for $z<0.04$);
the K-S test shows that the two distributions are significantly different
(a probability level 10$^{-5}$).
As such, the observed lower AGN fraction in LSBGs than that in HSBGs
in our study  is likely a manifestation of the postulation that
LSBGs are systematically smaller in size compared to HSBGs,
if the proposed AGN dependence on the physical scale of galaxies is the case.
However, these postulations need further confirmation
in future studies. 

\begin{figure}
\includegraphics[width=0.8\hsize,height=0.7\hsize]{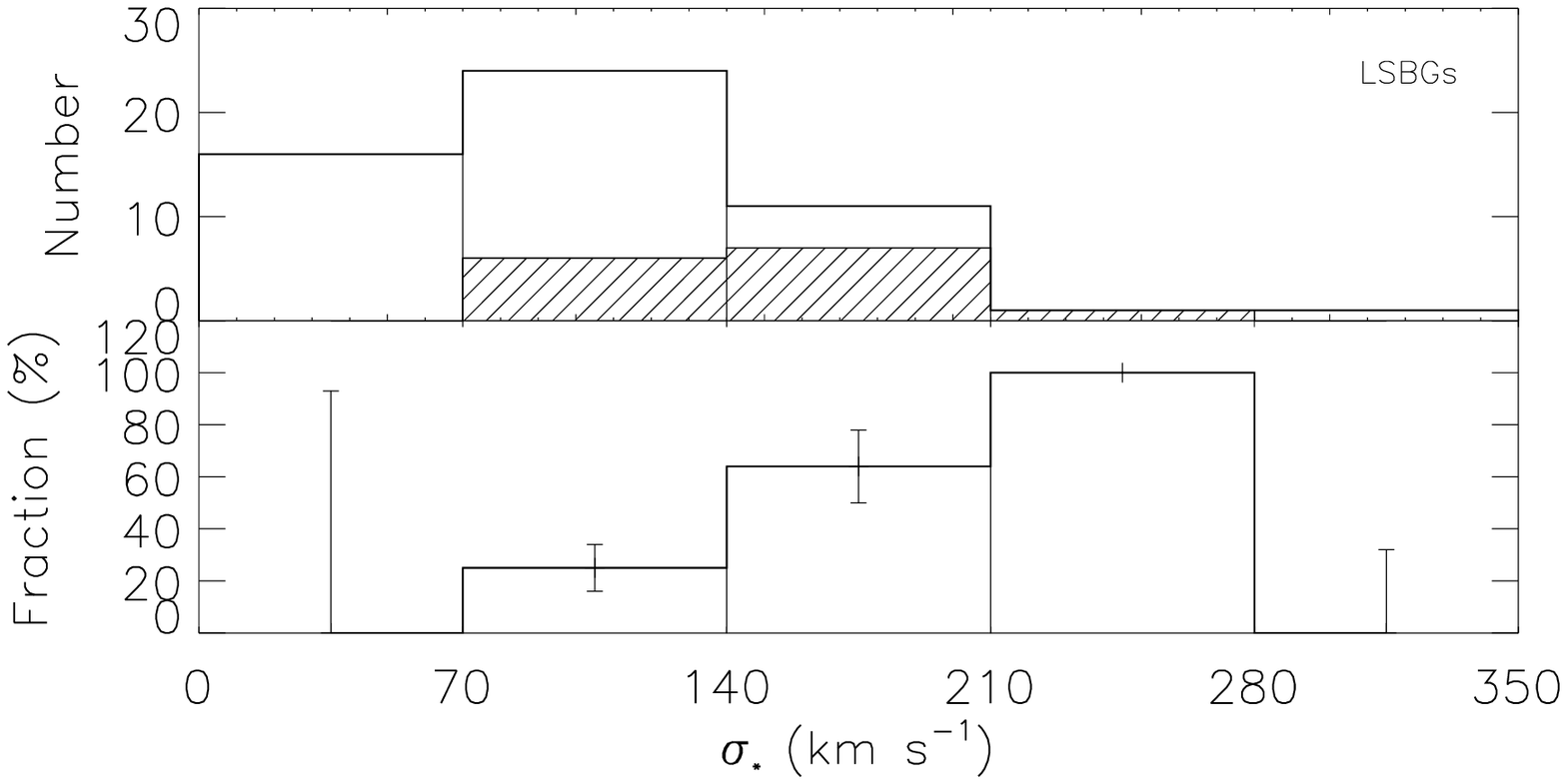}
\includegraphics[width=0.8\hsize,height=0.7\hsize]{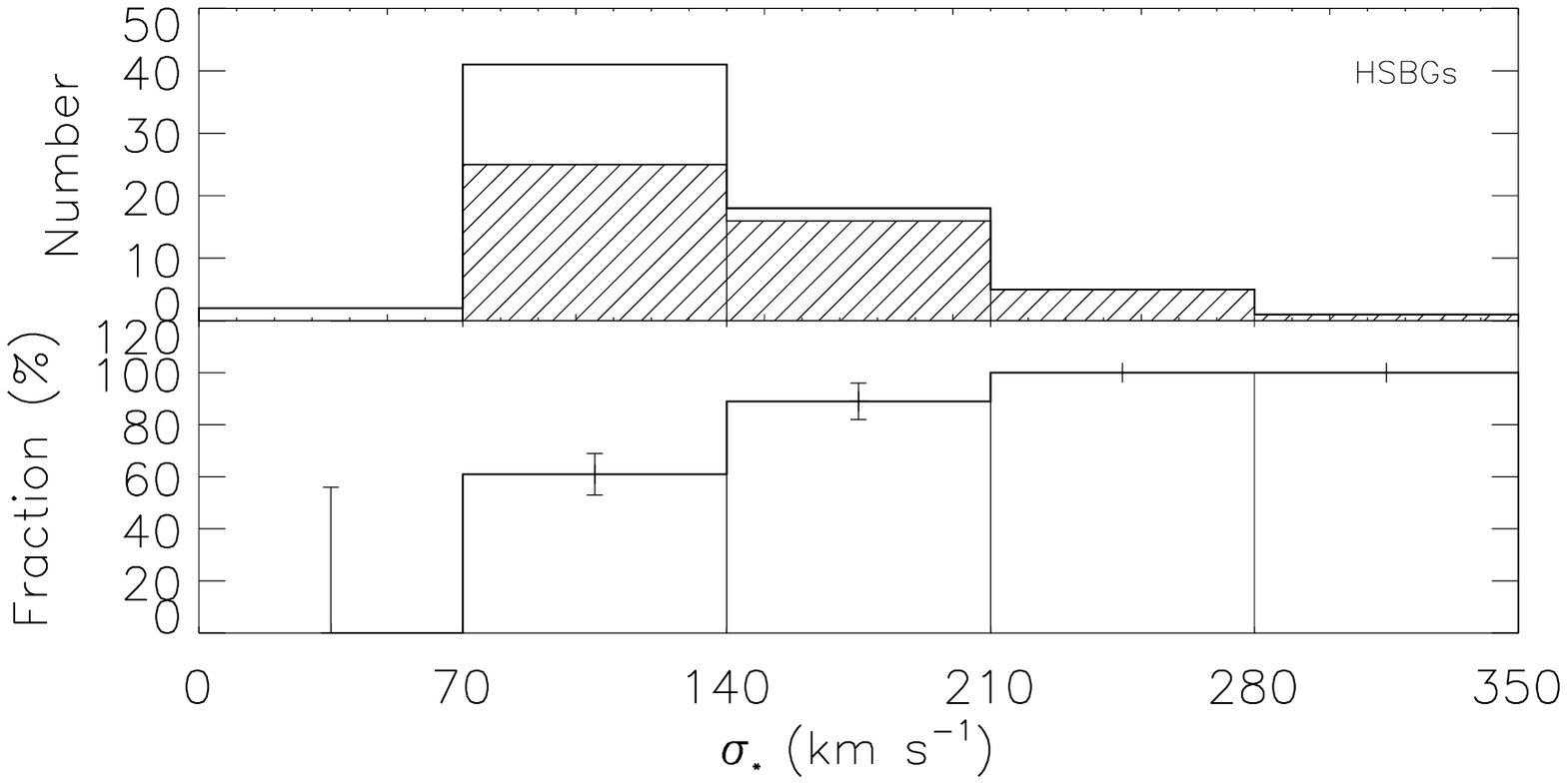}
\caption{Distributions of emission-line galaxies (top panel) and the 
AGN fraction (bottom panel) as a function of stellar velocity dispersion 
$\sigma_\ast$ of the LSBG sample and its comparing HSBG sample. The hatched histogram denotes AGN.}
\label{agn_frac_vdisp}
\end{figure}

We also test the possible dependence of AGN fraction on galactic bulge properties.
It has been found that stellar velocity dispersion $\sigma_*$
is well correlated with the bulge mass and luminosity,
and is thus a tracer of the gravitational potential of the bulge.
Given the spectral resolution of SDSS data,
the minimum values of $\sigma_*$ can be measured are around 70\,\kmpse.
For those objects without measurable  $\sigma_*$, 
we assume their  $\sigma_*$  to be less than  70\,\kmpse.
Figure\,\ref{agn_frac_vdisp} shows the distribution of  $\sigma_*$ 
for the APM-SDSS LSBG sample and for their AGNs,
as well as the AGN fraction.
It can be seen clearly that the AGN fraction increases in higher
 $\sigma_*$ bins.
Similar trend is also found in the HSBG comparing sample,
which is a known property that AGNs tend to be found in early
type galaxies, where bulges are dominant.
Our result in Section\,3 shows that even for the same Hubble type
(Sa--Sc),
LSBGs have a lower AGN fraction than HSBGs. 
In the context of the AGN fraction dependence on bulge properties,
this may imply that within a given morphological type,
LSBGs have either relatively smaller bulge mass or lower
fraction of galaxies with bulges, compared to HSBGs. 
Again, this postulation needs confirmation by future studies.
The dependence of AGN fraction on both the size of galaxies
and galactic bulge property is not surprising,
since galaxy sizes and $\sigma_*$ are strongly coupled in the
samples studied here.
However, we consider the latter relation is perhaps more fundamental,
since it has becoming known that the growth of galactic bulges
and the growth of central black hole are somehow related \citep{b6,b8}.

\subsection{Black hole growth in LSBGs}

  The broad H$\alpha$ emission lines are well detected in 3 LSBGs.
We estimate the masses of central black holes using the 
linewidth-luminosity mass scaling relation given in \citet{GreeneHo07}.
The values are 2.8 $\times$ 10$^{6}$ M$_\odot$ for SDSS J011448.7$-$002946.1, 
2.0 $\times$ 10$^{7}$ M$_\odot$ for SDSS J122912.9$+$004903.7 and 3.4 $\times$ 10$^{6}$ M$_\odot$ for SDSS J231815.7$+$001540.2 respectively.
For the two AGNs, SDSS J011448.7$-$002946.1 and SDSS J122912.9$+$004903.7,
broad components of H$\beta$ are also detected with reliable confidence.
It is claimed by \citet{b61} that the absolute uncertainties 
in masses estimated by these mass scaling relationships
are about a factor of 4.
According to the scaling relationships calibrated by 
\citet{b61} based upon the broad H$\beta$ luminosity,
black hole masses are estimated to be 3.6 $\times$ 10$^{6}$ M$_\odot$ for SDSS J011448.7$-$002946.1
and 4.4 $\times$ 10$^{7}$ M$_\odot$ for SDSS J122912.9$+$004903.7 are found, respectively.

\begin{figure}
\includegraphics[]{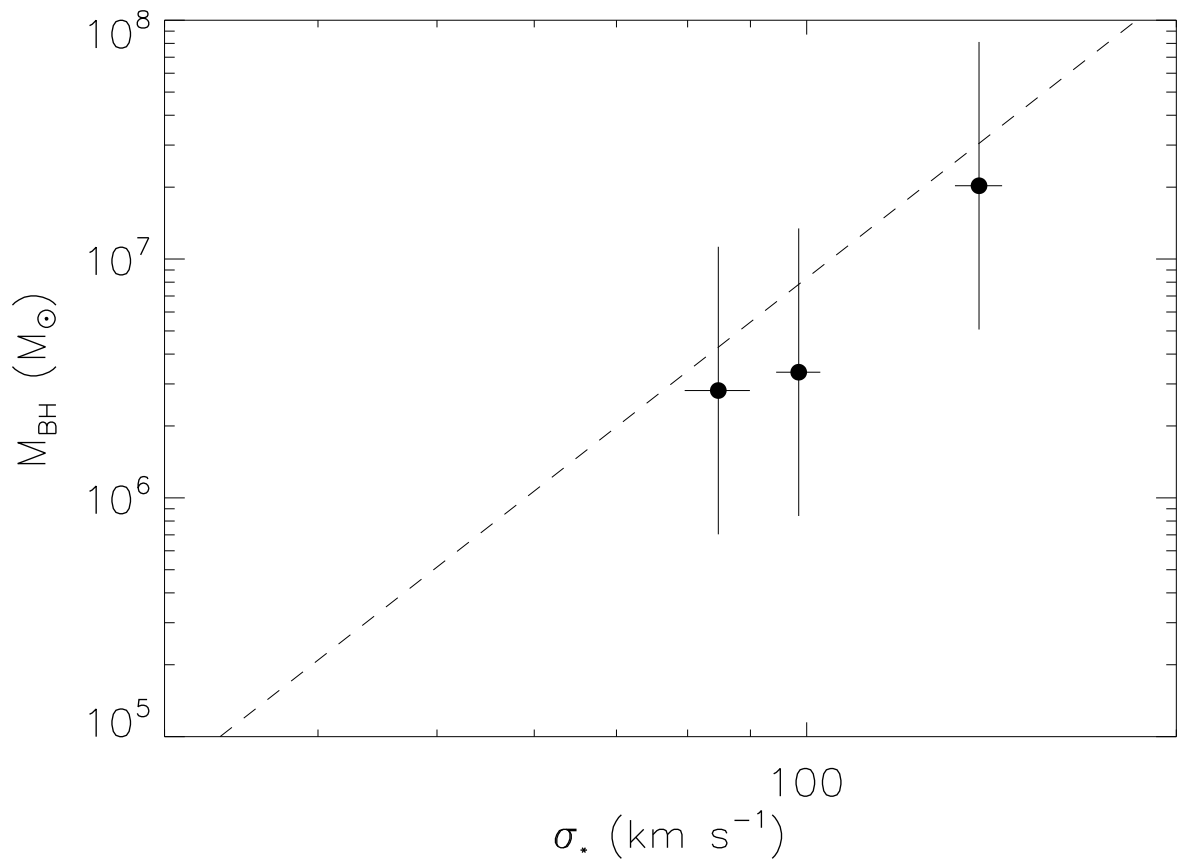}
\caption{Relation between the stellar velocity dispersion and the 
black hole mass for 3 broad line AGNs in LSBGs. 
The black hole masses are uncertain by a factor of 4. 
The dashed line is the relation given by \citet{b49} for local normal galaxies and AGNs.}
\label{mbh_sigma_star}
\end{figure}

For these 3 AGN with central black hole mass estimates, 
the stellar velocity dispersion  $\sigma_*$ can also be measured in our
spectral analysis.
We are therefore able to, for the first time,  test the relationship between
 black hole mass and stellar velocity dispersion (M$-\sigma_*$) relation
for LSBGs,  though the number of objects is very small.
The result is shown in Figure\,\ref{mbh_sigma_star}.
It can be seen that, for LSBG, the 
observed data are broadly consistent within errors with the
known M$-\sigma_*$ established based on local normal galaxies and AGNs
(Tremaine et al. 2002, dashed line).
However, a statistically meaningful result has to await 
the availability of a larger sample with both black hole mass and $\sigma_*$ measurements.


\section{Conclusion}
We have performed detailed spectral analysis of 194 LSBGs
from the Impey et al. (1996) APM LSBG sample  which have
been observed spectroscopically by the SDSS DR5.
Our work improves upon previous spectroscopic studies on LSBGs
with high and homogeneous quality SDSS spectra and elaborate
spectral analysis, which includes subtraction of host galaxy starlight
spectra, and  deblending narrow and broad components of
the Balmer lines and doublets.
These improvements allow us to carry out, for the first time,
reliable spectral classification of nuclear activities of LSBGs
based on the  BPT classification schemes in a rigorous way.
We also identified three secured broad line AGNs in LSBGs, and clarified
a few spurious ones claimed in previous work.

We found that, the majority (68\%) of the LSBGs are emission line galaxies.
The most abundant class is star-forming galaxies (52\%) characteristic of HII
emission line spectra.
Within the APM-SDSS sample, about 15\% LSBGs show emission line spectra
characteristic of AGNs, including Seyfert galaxies, LINERs, and
galactic nuclei composed of an AGN and central star-forming region.
Such a fraction of AGNs is significantly lower than that found in local
normal galaxies.
This result holds true even within each morphological type from Sa to Sc.
Our results do not support the high AGN fractions in LSBGs suggested in some
of the previous studies \citep{sprayberry,b21},
but is qualitatively consistent with that found by Impey01.
We found that the fraction of AGNs depends strongly  both on the
stellar velocity dispersion of the bulges and on the physical size of
the galaxies, the two of which are coupled with each other.
We interpret the low AGN fraction in LSBGs in terms of the postulation
that, compared to HSBGs,  LSBGs possess relatively lower bulge masses,
or a lower fraction of galaxies with bulges,
 even within the same morphological type.
This hypothesis can be tested in further observations in the future.

Compared to AGNs of HSBGs, AGNs in LSBGs tend to have relatively lower
electron density and lower dust extinction in the NLR.
As in HSBGs, the width of narrow emission lines in active nuclei
of LSBGs is a good tracer of the stellar velocity dispersion in the galactic bulge.
The black hole masses of the 3 broad line AGNs in LSBGs
are broadly consistent within errors with the well known M-$\sigma_\ast$ relation
found for nearby galaxies and AGNs.

\section{Acknowledgements} 
This work is supported by the
National Natural Science Foundation of China NSF-10533050 and 
NSF-10373004. 
This work has made use of the data products of the SDSS.
We thank the anonymous referee for reviewing this paper.
We also thank Jianguo Wang and Hongyan Zhou for their checking the SDSS spectra and valuable suggestion. 
We gratefully thank Poon Helen for her careful
English correction.

\appendix
\section{SDSS starlight spectral modeling}

The spectra are first corrected for Galactic extinction using the extinction
map of Schlegel et al.\ (1998) and
the reddening curve of Fitzpatrick (1999), and transformed into the
rest frame using the redshift provided by the SDSS pipeline.
Then host-galaxy starlight and AGN continuum,
as well as the optical \feii emission complex are modeled as

\begin{equation}\label{eq1}
S(\lambda)=A_{host}(E_{B-V}^{host},\lambda)~A(\lambda)
+A_{nucleus}(E_{B-V}^{nucleus},\lambda)~
[bB(\lambda)+ c_{\rm b} C_{\rm b}(\lambda) + c_{\rm n} C_{\rm n}(\lambda) ]
\end{equation}
where $S(\lambda)$ is the observed spectrum.
$A(\lambda)=\sum_{i=1}^6 a_{i}~IC_{i}(\lambda,\sigma_{*})$
represents the starlight component modeled by our 6 synthesized
galaxy templates, which had been built up from the spectral template
library of Simple Stellar Populations (SSPs) of
\citet{b33} using our new method based on the Ensemble
Learning Independent Component Analysis (EL-ICA) algorithm.
The details of the galaxy
templates and their applications are presented in \citep{b55}.
$A(\lambda)$ was broadened by convolving with a Gaussian of width
$\sigma_{*}$  to match the stellar velocity dispersion of the host galaxy.
The un-reddened nuclear continuum is assumed to be
$B(\lambda)=\lambda^{-1.7}$ as given in \citet{fra96}.
We modeled the optical \feii emission, both broad and narrow,
using the spectral data of the \feii
multiplets for I\,Zw\,I in the $\lambda\lambda$\,3535--7530\AA\
range provided by \citet{b35}[Table\,A1,A2]{veron04}.
We assume that the  broad F$_{e}$II lines ($C_{\rm b}$ in Eq.1)
have the same profile as the
broad \hb line, and the narrow \feii lines ($C_{\rm n}$),
 both permitted and forbidden, have the same profile as the that of the
narrow \hb component, or of \oiiie$\lambda5007$ if \hb is weak.
$A_{host}(E_{B-V}^{host},\lambda)$ and
$A_{nucleus}(E_{B-V}^{nucleus},\lambda)$ are the color excesses due
to possible extinction of the host galaxy and the nuclear region,
respectively, assuming the extinction curve for the Small Magellanic
Cloud of \citet{pei92}. The fitting is performed by minimizing
the $\chi^{2}$ with $E_{B-V}^{host}$, $E_{B-V}^{nucleus}$, $a_{i}$,
$\sigma_{*}$, $b$, $c_{\rm b}$ and $c_{\rm n}$ being free parameters.
To account for possible error of the redshifts provided by the SDSS pipeline,
in practice we loop possible redshifts near the SDSS redshift, 
spaced every 5 \kmps.
We fit and subtracted the above model from the SDSS spectra.

\clearpage \tablenum{1}
\begin{deluxetable}{llllllllllll}
\tabletypesize{\scriptsize}
\rotate
\tablecaption{Spectral parameters and classification of the 131 emission-line LSBGs}
\tablewidth{0pt}
\tablehead{
\colhead{Name} & 
\colhead {Z} & 
\colhead{[OII]\tablenotemark{a}} & 
\colhead{H$\beta$\tablenotemark{a}} & 
\colhead{[O III]\tablenotemark{a}} & 
\colhead{[OI]\tablenotemark{a}} & 
\colhead{H$\alpha$\tablenotemark{a}} & 
\colhead{[NII]\tablenotemark{a}} & 
\colhead{[SII] \tablenotemark{a}} & 
\colhead{FWHM[NII]\tablenotemark{b}} & 
\colhead{$\sigma_\ast$\tablenotemark{c}} & 
\colhead{Type\tablenotemark{d}} \\

\colhead{} & 
\colhead{} & 
\colhead{$\lambda$3727} & 
\colhead{narrow} & 
\colhead{$\lambda$5007} & 
\colhead{$\lambda$6300} & 
\colhead{narrow} & 
\colhead{ $\lambda$6583} & 
\colhead{$\lambda$6717/$\lambda$6731} & 
\colhead{km s$^{-1}$} & 
\colhead{km s$^{-1}$} & 
\colhead{} }
\startdata
001455.1$+$001508.3 & 0.039 & 289.6 & 400.6 & 122.2 & 61.1 & 1859.1 & 1062.5 & 
282.2/261.8 & 226.7 &       111.9 & composite\\
001558.2$-$001812.6 & 0.039 & *** & 43.6 & 89.1 & 46.0 & 163.3 & 129.0 & 88.0/
68.1 & 299.5 &       170.3 & LINER\\
001930.7$-$003606.3 & 0.033 & *** & 223.6 & 312.3 & 16.7 & 623.7 & 130.6 & 133.1
/93.2 & 172.2 &  ***  & hii\\
002149.8$-$001929.4 & 0.044 & *** & 4.1 & 6.0 & *** & 12.4 & 3.5 & 4.7/3.1 & 
163.6 &  ***  & hii\\
002534.4$+$005048.6 & 0.018 & *** & 36.3 & 15.9 & 7.0 & 114.0 & 37.9 & 30.9/21.9
 & 161.2 &  ***  & hii\\
003143.3$+$005402.8 & 0.018 & 75.9 & 12.7 & 13.4 & *** & 43.9 & 12.4 & 13.2/12.7
 & 169.6 &  ***  & hii\\
005042.7$+$002558.3 & 0.068 & 71.8 & 30.1 & 98.2 & 20.0 & 147.8 & 130.6 & 41.8/
38.2 & 424.0 &       163.4 & S2\\
005257.2$+$002317.4 & 0.035 & *** & 48.4 & 20.4 & 6.8 & 172.4 & 54.8 & 44.1/28.2
 & 166.7 &  ***  & hii\\
005257.8$+$002207.6 & 0.034 & *** & 1080.2 & 3204.9 & 81.7 & 4067.3 & 609.2 & 
424.7/274.9 & 179.4 &       147.7 & hii\\
005342.7$-$010506.6 & 0.047 & *** & 132.5 & 1179.5 & 143.7 & 349.7 & 594.8 & 8.1
/8.1 & 318.9 &       140.4 & LINER\\
005509.0$-$010247.3 & 0.046 & 41.8 & 32.8 & 46.5 & *** & 65.1 & 77.5 & 29.3/21.2
 & 367.5 &       170.6 & LINER\\
005848.9$+$003514.0 & 0.018 & *** & 163.2 & 53.3 & 15.2 & 670.5 & 287.9 & 107.1/
93.1 & 196.0 &  ***  & hii\\
005855.5$+$010017.4 & 0.018 & 182.5 & 713.7 & 3600.7 & 31.9 & 2469.7 & 65.8 & 
159.3/114.4 & 155.1 &       438.1 & hii\\
010550.2$+$001432.2 & 0.048 & *** & 151.6 & 40.8 & 18.3 & 617.6 & 203.4 & 93.8/
71.2 & 184.8 &       76.2 & hii\\
011050.8$+$001153.3 & 0.018 & *** & 93.9 & 19.4 & 5.0 & 409.4 & 149.5 & 63.9/
45.1 & 165.0 &  ***  & hii\\
011244.8$+$003935.1 & 0.065 & 35.9 & 38.1 & 15.1 & 7.6 & 139.2 & 53.8 & 30.9/
24.2 & 175.1 &       96.0 & hii\\
011310.0$+$005012.1 & 0.033 & 64.9 & 11.4 & 6.4 & *** & 42.1 & 11.1 & 13.4/10.3 
& 160.8 &  ***  & hii\\
011448.7$-$002946.1 & 0.034 & *** & 346.6 & 748.1 & 46.8 & 1478.3 & 704.8 & 
203.8/135.5 & 259.7 &       149.7 & S1\\
012121.3$+$000525.5 & 0.013 & *** & 31.4 & 24.5 & 13.7 & 65.6 & 17.0 & 27.7/14.9
 & 184.5 &       84.1 & hii\\
012539.7$+$011041.2 & 0.020 & *** & 53.3 & 12.1 & 11.9 & 207.1 & 81.6 & 39.3/
30.5 & 166.4 &  ***  & hii\\
021532.0$-$001727.4 & 0.026 & 63.0 & 16.6 & 9.4 & *** & 46.9 & 11.5 & 12.4/8.5 &
 169.5 &  ***  & hii\\
022606.7$-$001954.9 & 0.021 & 38.6 & 22.5 & 39.7 & *** & 91.9 & 89.3 & 36.9/26.8
 & 281.4 &       112.1 & LINER\\
022933.9$+$002223.3 & 0.021 & *** & 14.0 & 25.8 & *** & 31.9 & 6.6 & 11.7/8.7 & 
189.1 &  ***  & hii\\
023143.4$+$001736.5 & 0.021 & *** & 18.5 & 8.4 & 7.4 & 54.8 & 16.3 & 17.2/13.1 &
 171.1 &  ***  & hii\\
023238.1$+$003539.3 & 0.022 & 59.2 & 1018.5 & 3510.1 & 68.7 & 2862.3 & 215.1 & 
325.9/218.8 & 170.5 &       98.1 & hii\\
023239.3$+$003702.4 & 0.021 & 29.3 & 172.7 & 86.8 & 29.8 & 669.1 & 269.2 & 121.1
/75.1 & 189.0 &  ***  & hii\\
023601.0$+$002512.8 & 0.009 & *** & 14.8 & 9.8 & *** & 36.3 & 13.4 & 15.0/9.8 & 
162.4 &  ***  & hii\\
024056.6$+$001445.6 & 0.027 & *** & 11.3 & 10.1 & 10.2 & 29.0 & 9.1 & 9.8/5.4 & 
179.3 &  ***  & hii\\
024227.3$+$010214.4 & 0.046 & *** & 35.4 & 17.2 & 5.8 & 152.9 & 70.2 & 30.3/24.2
 & 244.4 &       100.0 & hii\\
024547.6$-$001427.1 & 0.023 & *** & 8.4 & 7.5 & *** & 18.4 & 9.2 & 9.0/5.7 & 
166.6 &  ***  & composite\\
024631.1$-$002158.5 & 0.051 & *** & 12.8 & 4.9 & *** & 54.5 & 20.4 & 12.6/7.4 & 
159.3 &  ***  & hii\\
031859.4$+$011347.5 & 0.038 & 198.3 & 6.5 & 9.4 & *** & 27.4 & 5.4 & 9.0/5.4 & 
142.3 &  ***  & hii\\
032910.8$-$010246.3 & 0.036 & *** & 7.9 & 4.7 & *** & 19.3 & 3.8 & 5.2/3.5 & 
152.3 &  ***  & hii\\
033507.2$-$005237.9 & 0.038 & *** & 5.9 & 18.0 & *** & 25.8 & 9.5 & 8.3/3.1 & 
166.6 &  ***  & composite\\
034907.9$+$010943.4 & 0.014 & 70.8 & 923.3 & 2489.0 & 40.7 & 2931.6 & 280.3 & 
315.4/224.2 & 163.5 &  ***  & hii\\
035326.2$+$005030.5 & 0.038 & 100.4 & 33.4 & 32.2 & *** & 122.4 & 73.1 & 25.2/
19.9 & 192.0 &       113.1 & composite\\
091613.8$+$004202.3 & 0.038 & 43.7 & 320.2 & 227.7 & 44.4 & 1386.2 & 668.9 & 
225.7/183.5 & 285.1 &       121.7 & composite\\
091745.3$+$010319.5 & 0.027 & 36.8 & 37.1 & 22.2 & 7.9 & 140.3 & 39.3 & 34.4/
22.2 & 165.8 &       97.5 & hii\\
091955.2$-$003528.9 & 0.029 & *** & 10.2 & 15.9 & *** & 49.7 & 8.1 & 15.3/10.8 &
 162.7 &  ***  & hii\\
092346.4$+$024510.7 & 0.017 & *** & 57.3 & 89.6 & 8.9 & 194.4 & 25.7 & 34.9/26.7
 & 173.4 &  ***  & hii\\
093223.1$+$023251.4 & 0.017 & 617.8 & 15.5 & 30.7 & *** & 52.4 & 6.0 & 10.8/7.9 
& 167.1 &  ***  & hii\\
095959.3$+$003817.6 & 0.033 & 33.9 & 67.1 & 100.3 & 13.6 & 260.2 & 43.3 & 61.2/
43.3 & 170.4 &  ***  & hii\\
100440.8$+$002211.6 & 0.045 & 106.5 & 19.8 & 28.7 & *** & 52.2 & 67.2 & 43.1/
22.3 & 339.1 &       124.8 & LINER\\
101105.2$+$011326.7 & 0.033 & *** & 63.5 & 19.1 & *** & 320.3 & 105.7 & 39.8/
32.1 & 224.5 &       98.0 & hii\\
101117.9$+$002632.6 & 0.012 & 42.4 & 201.9 & 34.5 & 15.0 & 841.1 & 281.6 & 93.5/
71.6 & 173.5 &  ***  & hii\\
101853.9$+$021439.2 & 0.046 & 817.8 & 36.2 & 30.2 & *** & 127.3 & 30.7 & 32.6/
19.3 & 161.4 &  ***  & hii\\
102533.3$+$021903.6 & 0.064 & 459.7 & 50.2 & 29.4 & *** & 186.4 & 82.1 & 26.8/
19.0 & 267.2 &       99.3 & hii\\
103135.1$-$005624.5 & 0.033 & *** & 15.6 & 12.6 & *** & 43.4 & 16.7 & 15.1/10.7 
& 191.0 &  ***  & hii\\
103226.9$+$023318.0 & 0.022 & 21.0 & 20.6 & 25.4 & 5.9 & 82.9 & 19.4 & 27.7/17.8
 & 171.4 &  ***  & hii\\
103321.0$+$023658.1 & 0.029 & 316.9 & 329.1 & 64.6 & 39.6 & 1675.2 & 594.3 & 
229.8/182.1 & 213.8 &       71.7 & hii\\
103723.6$+$021845.5 & 0.040 & 34.7 & 112.0 & 250.2 & 127.6 & 411.7 & 400.0 & 
256.4/205.3 & 390.0 &       209.8 & LINER\\
103725.7$+$020443.1 & 0.074 & 26.5 & 20.5 & 16.9 & 6.5 & 104.5 & 52.9 & 31.2/
22.5 & 358.9 &       155.7 & composite\\
103727.7$+$020521.9 & 0.073 & *** & 9.0 & 39.2 & 42.9 & 73.9 & 111.4 & 57.4/39.7
 & 485.0 &       196.0 & LINER\\
103825.5$-$000104.2 & 0.019 & *** & 14.3 & 12.0 & 3.2 & 50.2 & 10.6 & 19.4/12.2 
& 170.7 &  ***  & hii\\
104405.9$-$005744.5 & 0.027 & 40.6 & 7.4 & 8.6 & *** & 25.5 & 4.3 & 7.4/5.2 & 
166.4 &  ***  & hii\\
104501.2$+$013855.3 & 0.028 & 9.0 & 13.1 & 10.9 & *** & 36.6 & 8.6 & 10.6/8.6 & 
175.5 &  ***  & hii\\
104509.1$+$000433.4 & 0.094 & *** & 12.1 & 18.5 & 15.1 & 19.9 & 23.3 & 16.8/10.1
 & 345.2 &       180.6 & LINER\\
104614.9$+$000300.9 & 0.047 & 280.2 & 494.9 & 355.0 & 63.1 & 1817.9 & 539.2 & 
357.4/271.5 & 192.2 &       101.6 & hii\\
104634.3$+$014627.8 & 0.022 & *** & 10.4 & 8.1 & *** & 26.0 & 3.8 & 6.4/6.1 & 
162.3 &  ***  & hii\\
104819.9$-$000119.1 & 0.039 & 24.4 & 10.3 & 4.2 & 3.3 & 52.3 & 22.7 & 12.4/8.2 &
 172.9 &       61.9 & hii\\
105318.6$+$023733.7 & 0.003 & 85.3 & 185.5 & 371.0 & 13.5 & 556.8 & 17.7 & 56.4/
36.2 & 154.2 &  ***  & hii\\
110700.5$+$001022.3 & 0.038 & *** & 13.1 & 17.2 & *** & 33.8 & 6.4 & 7.4/5.6 & 
176.5 &  ***  & hii\\
111549.4$+$005137.5 & 0.046 & 36.0 & 161.2 & 219.8 & 290.0 & 972.7 & 815.5 & 
641.4/473.4 & 554.6 &       187.7 & LINER\\
111849.6$+$003709.5 & 0.025 & 109.6 & 123.3 & 33.8 & 15.7 & 462.7 & 162.3 & 81.6
/56.1 & 188.0 &  ***  & hii\\
112409.2$+$004202.0 & 0.026 & 739.0 & 58.6 & 18.4 & 5.7 & 246.4 & 89.2 & 40.5/
27.6 & 177.3 &  ***  & hii\\
112712.2$-$005940.8 & 0.003 & 390.8 & 74.3 & 109.1 & 9.3 & 200.4 & 16.4 & 46.8/
35.4 & 158.4 &  ***  & hii\\
112718.8$-$010335.5 & 0.041 & 255.7 & 7.3 & 5.4 & *** & 21.0 & 5.9 & 6.8/4.3 & 
159.5 &  ***  & hii\\
113245.4$-$004427.8 & 0.022 & *** & 34.5 & 24.7 & 4.6 & 134.1 & 29.5 & 39.1/27.0
 & 155.7 &  ***  & hii\\
113505.0$+$023304.1 & 0.017 & 28.5 & 207.4 & 424.9 & 17.0 & 706.4 & 97.2 & 107.4
/78.7 & 157.0 &       83.8 & hii\\
115336.9$+$020957.8 & 0.040 & *** & 9.6 & 21.2 & *** & 39.0 & 4.4 & 10.4/5.0 & 
164.0 &  ***  & hii\\
115342.8$-$013935.6 & 0.011 & 71.3 & 33.3 & 56.2 & *** & 93.0 & 15.0 & 24.2/13.9
 & 166.5 &  ***  & hii\\
115412.1$-$002856.7 & 0.006 & 132.0 & 9.8 & 24.3 & *** & 33.1 & 2.6 & 4.5/3.0 & 
159.6 &  ***  & hii\\
115801.1$-$021038.3 & 0.082 & 90.4 & 13.4 & 20.0 & *** & 22.3 & 30.9 & 9.6/22.7 
& 409.8 &       187.3 & LINER\\
115924.6$+$012602.6 & 0.047 & *** & 25.8 & 5.3 & *** & 73.6 & 29.0 & 12.2/11.4 &
 207.0 &       64.3 & hii\\
120804.0$+$004151.3 & 0.020 & 38.7 & 103.6 & 100.1 & 9.5 & 301.9 & 66.0 & 57.6/
38.3 & 160.0 &  ***  & hii\\
120806.2$-$023156.0 & 0.026 & *** & 14.2 & 13.9 & *** & 39.0 & 14.5 & 12.7/9.3 &
 153.3 &  ***  & hii\\
121159.5$+$012100.1 & 0.021 & *** & 82.5 & 56.3 & 11.1 & 273.7 & 67.6 & 70.9/
49.5 & 168.6 &       67.0 & hii\\
121203.3$-$003621.7 & 0.035 & 22.9 & 708.0 & 1899.6 & 46.7 & 2308.2 & 223.9 & 
266.7/188.8 & 192.0 &       336.1 & hii\\
121248.3$-$024328.6 & 0.038 & *** & 247.9 & 169.9 & 89.6 & 1228.5 & 670.2 & 
331.6/293.6 & 379.4 &       169.8 & composite\\
121431.0$+$021000.7 & 0.074 & *** & 18.9 & 7.4 & *** & 73.9 & 26.2 & 9.3/8.6 & 
189.5 &  ***  & hii\\
121604.5$+$011049.6 & 0.050 & *** & 237.1 & 74.6 & 47.7 & 1175.7 & 488.3 & 186.1
/148.1 & 260.9 &       98.7 & hii\\
121638.7$-$012706.7 & 0.003 & 1975.4 & 29.4 & 45.7 & 6.0 & 84.4 & 4.7 & 8.9/6.3 
& 177.9 &  ***  & hii\\
122033.8$+$004717.6 & 0.007 & *** & 81.3 & 96.8 & 11.4 & 249.9 & 38.7 & 52.4/
38.7 & 160.2 &  ***  & hii\\
122404.9$+$011123.3 & 0.024 & 1001.8 & 413.2 & 75.1 & 24.8 & 1839.0 & 636.5 & 
228.4/185.3 & 227.7 &       107.4 & hii\\
122610.0$-$010923.1 & 0.042 & *** & 44.3 & 25.6 & *** & 151.2 & 68.8 & 36.9/26.0
 & 174.2 &  ***  & composite\\
122801.7$+$013629.5 & 0.077 & *** & 57.6 & 18.3 & 9.6 & 252.2 & 86.8 & 39.1/30.2
 & 175.6 &  ***  & hii\\
122912.9$+$004903.7 & 0.079 & *** & 30.7 & 200.7 & 11.5 & 97.1 & 91.3 & 33.1/
24.5 & 298.8 &       168.2 & S1\\
122921.6$+$010325.0 & 0.023 & 23.2 & 166.0 & 32.9 & 17.3 & 683.7 & 212.1 & 121.6
/90.9 & 171.9 &       67.4 & hii\\
125323.8$-$002523.5 & 0.048 & *** & 14.7 & 6.3 & 11.8 & 100.7 & 40.9 & 23.9/13.6
 & 172.4 &  ***  & hii\\
130058.7$-$000139.1 & 0.004 & *** & 1330.1 & 479.9 & 75.3 & 5002.2 & 1602.0 & 
821.0/615.2 & 181.1 &       74.9 & hii\\
130243.5$+$003949.9 & 0.041 & *** & 90.6 & 22.9 & 13.2 & 417.6 & 202.1 & 62.4/
48.9 & 205.5 &       95.2 & hii\\
130316.0$+$012807.1 & 0.041 & 49.4 & 39.3 & 13.2 & *** & 202.0 & 79.5 & 38.7/
30.8 & 197.0 &       93.8 & hii\\
130338.9$+$024335.2 & 0.071 & *** & 22.9 & 51.5 & *** & 57.4 & 44.0 & 36.6/21.9 
& 405.9 &       207.5 & S2\\
131004.7$+$005655.7 & 0.019 & *** & 8.2 & 4.9 & *** & 20.1 & 4.4 & 5.9/4.0 & 
198.4 &  ***  & hii\\
131809.3$+$001322.1 & 0.032 & *** & 34.8 & 30.9 & *** & 120.8 & 36.2 & 37.5/24.5
 & 164.0 &  ***  & hii\\
132340.7$+$012142.5 & 0.057 & *** & 46.8 & 57.7 & 29.8 & 296.5 & 204.4 & 76.0/
58.4 & 271.3 &       132.6 & composite\\
132955.8$+$013238.6 & 0.004 & *** & 79.7 & 136.7 & 9.7 & 251.5 & 19.0 & 47.2/
32.0 & 163.1 &  ***  & hii\\
133032.0$-$003613.5 & 0.054 & *** & 290.5 & 856.5 & 96.8 & 1553.5 & 995.0 & 
334.8/270.8 & 299.5 &       158.5 & S2\\
133305.3$-$010208.9 & 0.012 & 34.5 & 181.7 & 287.2 & *** & 835.4 & 1194.3 & 
616.3/473.9 & 378.0 &       133.3 & LINER\\
135658.2$-$010606.1 & 0.031 & *** & 30.0 & 24.5 & *** & 93.5 & 21.5 & 16.5/11.1 
& 173.9 &  ***  & hii\\
140043.0$-$003020.5 & 0.012 & 48.8 & 104.2 & 164.3 & 8.9 & 285.2 & 43.7 & 49.8/
37.0 & 157.2 &  ***  & hii\\
140126.7$-$024245.9 & 0.030 & 63.0 & 26.2 & 14.1 & *** & 113.3 & 29.1 & 27.9/
17.8 & 157.2 &       70.7 & hii\\
140127.5$+$015822.6 & 0.024 & *** & 21.6 & 30.9 & 6.6 & 82.1 & 14.9 & 22.7/19.0 
& 172.5 &  ***  & hii\\
140320.7$-$003259.8 & 0.025 & 56.3 & 489.0 & 78.8 & 43.1 & 2173.5 & 857.7 & 
300.1/226.6 & 198.5 &       78.7 & hii\\
140321.1$-$003256.9 & 0.025 & 2265.5 & 384.0 & 95.1 & 23.9 & 1498.4 & 485.7 & 
216.1/148.2 & 168.0 &       66.2 & hii\\
140831.6$-$000737.4 & 0.025 & 95.6 & 39.9 & 17.2 & 9.8 & 136.9 & 50.8 & 39.3/
24.0 & 171.3 &  ***  & hii\\
141130.1$+$004159.9 & 0.053 & 665.4 & 9.4 & 12.7 & *** & 39.3 & 8.3 & 12.7/6.0 &
 177.5 &  ***  & hii\\
141132.9$-$031311.1 & 0.030 & *** & 87.0 & 22.6 & *** & 386.5 & 133.4 & 52.4/
38.3 & 226.2 &       99.6 & hii\\
141140.7$-$010307.8 & 0.024 & 14.7 & 6.9 & 12.7 & *** & 36.0 & 4.7 & 9.9/6.2 & 
152.9 &  ***  & hii\\
143127.2$-$024009.0 & 0.024 & 33.0 & 25.6 & 12.6 & *** & 43.5 & 6.1 & 10.8/9.0 &
 177.6 &  ***  & hii\\
143421.4$+$013626.5 & 0.031 & *** & 26.2 & 16.8 & 2.5 & 70.1 & 15.8 & 19.9/10.1 
& 167.8 &  ***  & hii\\
143638.6$+$000702.6 & 0.030 & 13.0 & 14.3 & 6.6 & *** & 42.1 & 15.1 & 13.6/10.8 
& 177.4 &  ***  & hii\\
143846.3$+$010657.7 & 0.083 & *** & 20.5 & 62.8 & 15.5 & 151.2 & 118.6 & 70.1/
52.7 & 547.8 &       271.0 & S2\\
144245.9$-$002103.9 & 0.006 & 54.9 & 221.9 & 496.9 & 27.6 & 661.0 & 80.4 & 132.3
/89.8 & 170.4 &  ***  & hii\\
144500.2$+$012430.8 & 0.034 & *** & 22.1 & 14.1 & *** & 95.9 & 55.9 & 20.3/17.0 
& 212.4 &       88.8 & composite\\
144525.4$+$001404.3 & 0.038 & *** & 21.6 & 16.2 & 6.6 & 87.6 & 32.7 & 12.6/9.0 &
 178.7 &       70.4 & hii\\
144620.6$-$010520.0 & 0.029 & *** & 786.3 & 3185.1 & 59.9 & 2697.6 & 110.3 & 
269.5/188.7 & 158.6 &  ***  & hii\\
144702.4$-$022307.2 & 0.030 & *** & 8.8 & 5.1 & *** & 22.6 & 6.3 & 6.1/6.3 & 
164.0 &  ***  & hii\\
144856.4$-$004338.0 & 0.028 & *** & 25.3 & 30.2 & *** & 87.1 & 15.0 & 19.3/15.8 
& 168.2 &  ***  & hii\\
144902.6$+$022611.2 & 0.034 & *** & 11.2 & 8.2 & *** & 34.8 & 21.0 & 10.7/4.2 & 
182.0 &  ***  & composite\\
144923.4$-$013333.4 & 0.027 & *** & 19.9 & 23.3 & *** & 58.5 & 8.7 & 14.8/8.6 & 
163.7 &  ***  & hii\\
145325.3$+$021810.9 & 0.027 & *** & 5.7 & 10.3 & *** & 20.3 & 2.9 & 7.4/4.7 & 
146.4 &  ***  & hii\\
145423.9$-$010747.3 & 0.023 & *** & 19.4 & 22.6 & *** & 54.6 & 10.1 & 14.9/10.5 
& 172.6 &  ***  & hii\\
231221.0$-$010542.4 & 0.026 & *** & 13.3 & 14.5 & *** & 39.3 & 10.3 & 13.9/9.3 &
 165.5 &  ***  & hii\\
231501.6$+$000420.1 & 0.051 & *** & 23.8 & 39.0 & 33.1 & 77.2 & 88.8 & 45.6/42.6
 & 338.2 &       121.6 & LINER\\
231815.7$+$001540.2 & 0.030 & *** & 186.0 & 514.3 & 52.6 & 756.1 & 483.6 & 220.1
/178.5 & 274.9 &       124.5 & S1\\
231952.0$+$011305.0 & 0.030 & *** & 54.4 & 55.7 & 8.6 & 160.1 & 28.4 & 44.2/27.3
 & 169.6 &  ***  & hii\\
232021.2$-$001819.3 & 0.025 & *** & 14.5 & 15.4 & *** & 49.2 & 12.7 & 17.7/8.3 &
 162.7 &  ***  & hii\\
232151.8$-$004126.8 & 0.024 & *** & 53.9 & 15.1 & 6.6 & 244.9 & 77.7 & 47.4/35.3
 & 173.1 &       81.2 & hii\\
233646.8$+$003724.5 & 0.009 & 27.3 & 125.8 & 315.5 & 5.4 & 356.2 & 37.4 & 49.4/
34.1 & 155.0 &  ***  & hii\\
234422.2$+$000547.0 & 0.022 & *** & 17.8 & 4.6 & *** & 21.9 & 8.4 & 10.9/5.2 & 
172.9 &  ***  & hii\\
\enddata
\tablenotetext{a}{In units of $10^{-17}$ erg cm$^{-2}$ s$^{-1}$}
\tablenotetext{b}{The full width at half maximum (FWHM), in units of km s$^{-1}$}
\tablenotetext{c}{The measured stellar velocity dispersion of galactic bulge, in units of km s$^{-1}$}
\tablenotetext{d}{The classification of emission-line nuclear based on the narrow line fluxes ratios.}
\label{131_lsbgs}
\end{deluxetable}

\tablenum{2}
\begin{deluxetable}{lcccccccc}
\tabletypesize{\scriptsize}
\rotate
\tablecaption{Parameters of broad Balmer lines of 3 LSB AGNs with broad emission lines.}
\tablewidth{0pt}
\tablehead{
\colhead{Name} & 
\colhead{f[H$\beta$]\tablenotemark{e}} & 
\colhead{FWHM (H$\beta$)} &
\colhead{f[H$\alpha$]\tablenotemark{e}} & 
\colhead{FWHM (H$\alpha$)} & 
\colhead{$\sigma_\ast$} & 
\colhead{M$_{BH}$\tablenotemark{f}} \\

\colhead{SDSS} & 
\colhead{$\lambda$4861$^{broad}$} & 
\colhead{km s$^{-1}$} & 
\colhead{ $\lambda$6563$^{broad}$} & 
\colhead{km s$^{-1}$} & 
\colhead{km s$^{-1}$} &
\colhead{M$_\odot$} \\
}
\startdata
J011448.7$-$002946.1 & 743.4 &3195 & 2769.4 & 2418 & 84.7 &2.8 $\times$ $10^{6}$ \\
J122912.9$+$004903.7 & 298.9 & 7400&1192.6&4750 & 138.2 & 2.0 $\times$ $10^{7}$ \\
J231815.7$+$001540.2&***&***&1037.4&3756 & 98.5 & 3.4 $\times$ $10^{6}$ \\
\enddata
\tablenotetext{e}{In units of $10^{-17}$ erg cm$^{-2}$ s$^{-1}$}
\tablenotetext{f}{Estimated black hole masses based on equations from \citet{GreeneHo07}}
\label{3_broad_lsbgs}
\end{deluxetable}

\clearpage
\tablenum{3}
\begin{deluxetable}{lcccccccc}
\tabletypesize{\scriptsize}
\tablecaption{Properties of the LSBGs with an AGN.}
\tablewidth{0pt}
\tablehead{
\colhead{Name} & 
\colhead{Name} & 
\colhead{\mube\footnote{Central surface brightness in B band}} & 
\colhead{$\mu_{\rm e}$ \footnote{Surface brightness in B band at the effective radius}} & 
\colhead{$B_{\rm total}$ \footnote{Total apparent magnitude in Johnson B band}} &
\colhead{$log(M_{\rm HI})$ \footnote{Logarithm of the neutral hydrogen mass in solar masses}} & 
\colhead{$R_{\rm e}$ \footnote{Effective radius in arcseconds,
defined as the radius of a circular aperture that encloses one$-$half the total intensity received from the galaxy}} & 
\colhead{$M_{\rm B}$ \footnote{Absolute magnitude in B band}} & 
\colhead{Hubble Type\footnote{Morphological classification in the system of de Vaucouleurs et al. (1991).}}
}
\startdata
0012$-$0001  & J001455.1$+$001508.3 & 20.8 & 21.5 & 15.5 &  & 5.7 & $-$19.9 & Sb\\
0013$-$0034  &J001558.2$-$001812.6&20.3&23.3&14.5&10.18&18.4&$-$20.9&Sc\\
0048$+$0009  &J005042.7$+$002558.3&21.3&22.3&15.8& & 6.2&$ $&Sb\\
0051$-$0121  &J005342.7$-$010506.6&21.6&22.9&15.9& & 7.2&$ $&SBc\\
0052$-$0119  &J005509.0$-$010247.3&19.9&23.4&14.9& &11.9&$-$20.8&Sd\\
0112$-$0045  &J011448.7$-$002946.1&21.5&22.0&14.9& & 8.9&       &Interacting\\
0223$-$0033  &J022606.7$-$001954.9&22.1&23.3&14.5& 9.83&18.3&$-$19.5&Sc\\
0243$-$0027  &J024547.6$-$001427.1&22.6&24.2&17.0& & 8.4& &Sc\\
0332$-$0102  &J033507.2$-$005237.9&22.9&23.9&17.3& & 6.7& &Sc\\
0350$+$0041  &J035326.2$+$005030.5&21.9&23.5&16.1& 9.60& 9.7&$-$19.2&Sc\\
0913$+$0054  &J091613.8$+$004202.3&20.4&22.6&15.5& 9.68& 8.2&$-$19.8&Sm \\
1002$+$0036  &J100440.8$+$002211.6&20.4&23.0&15.8& & 6.5&$-$19.8&SBb\\
1034$+$0234  &J103723.6$+$021845.5&20.2&22.0&15.4& & 6.5&$-$20.0&Interacting\\
1034$+$0220xc&J103725.7$+$020443.1&21.6&22.7&17.8& & 3.1&$-$19.0&Sbc\\
1034$+$0220  &J103727.7$+$020521.9&21.0&23.8&16.1& & 8.8&$-$20.6&Sc\\
1042$+$0020  &J104509.1$+$000433.4&21.4&23.4&16.2& & 9.6&$-$21.2&Irr\\
1113$+$0107  &J111549.4$+$005137.5&21.4&24.5&16.2& &11.4&$-$19.6&Interacting\\
1155$-$0153  &J115801.1$-$021038.3&20.7&23.6&15.9& & 9.9&$-$21.1&Sc\\
1210$-$0226  &J121248.3$-$024328.6&20.6&22.5&15.5& & 6.9&$-$19.8&Interacting\\
1223$-$0052  &J122610.0$-$010923.1&22.5&22.8&17.3& & 4.4&  &Sa\\
1226$+$0105  &J122912.9$+$004903.7&20.9&23.9&15.7&10.26&12.9&$-$21.3&Sc\\
1301$+$0259  &J130338.9$+$024335.2&21.9&25.4&16.9& & 9.7&$-$19.7&Sc\\
1321$+$0137  &J132340.7$+$012142.5&21.5&22.3&15.9& 9.64& 6.7&$-$20.3&Sm\\
1327$-$0020  &J133032.0$-$003613.5&20.0&20.7&14.9& & 5.2&$-$21.2&Interacting\\
1330$-$0046  &J133305.3$-$010208.9&18.2&19.0&11.7& & 8.9&$-$21.1&SBb\\
1436$+$0119  &J143846.3$+$010657.7&21.2&23.1&16.2& & 7.5&$-$20.8&Sc\\
1442$+$0137  &J144500.2$+$012430.8&22.0&22.9&16.5& & 6.3&$-$18.5&Sc\\
1446$+$0238  &J144902.6$+$022611.2&21.8&23.4&16.0& 9.73& 9.5&$-$19.1&Sc\\
2312$-$0011  &J231501.6$+$000420.1&20.7&21.5&15.6& 9.98& 5.5&$-$20.4&Sb\\
2315$-$0000  &J231815.7$+$001540.2&20.8&23.0&15.0& 9.71&12.3&$-$19.8&Sc\\
\enddata
\tablenotetext{2}{Central surface brightness in B band $($mag arcsec$^{-2})$}
\tablenotetext{3}{Surface brightness in B band at the effective radius $($mag arcsec$^{-2})$}
\tablenotetext{4}{Total apparent magnitude in Johnson B band (mag)}
\tablenotetext{5}{Logarithm of the neutral hydrogen mass in solar masses (M$_\odot$)}
\tablenotetext{6}{Effective radius in arcseconds, defined as the radius of a circular 
aperture that encloses one$-$half the total intensity received from the galaxy (arcsec)}
\tablenotetext{7}{Absolute magnitude in B band (mag)}
\tablenotetext{8}{Morphological classification in the system of de Vaucouleurs et al. (1991).}
\label{impey_30_agn}
\end{deluxetable}

\tablenum{4}
\begin{deluxetable}{lcccccc}
\tabletypesize{\scriptsize}
\tablecaption{Statistics and the fraction of each 
type of active nuclear for the LSBG samples.}
\tablewidth{0pt}
\tablehead{
\colhead{Sample} & \colhead{Seyfert} & \colhead{LINER} & \colhead{Transition} & \colhead{AGN} &
\colhead{Star-forming} & \colhead{Emission-line LSBGs}
}
\startdata
LSBGs (all)  & 7 (4$\%$) & 12 (6$\%$) & 11 (6$\%$) & 30 (16$\%$) & 101 (52$\%$) & 131 (68$\%$) \\
LSBGs (\mube $\geq 22.0$) & 0 (0.3$\%$) & 1 (0.8$\%$) & 4 (3$\%$) & 5 (4$\%$) & 58 (5$\%$) & 63 (54$\%$) \\
LSBGs (\mube $< 22.0$) & 7 (9$\%$) & 11 (14$\%$) & 7 (9$\%$) & 25 (32$\%$) & 43 (54$\%$) & 68 (86$\%$) \\
\enddata
\label{frac_mu}
\end{deluxetable}

\tablenum{5}
\begin{deluxetable}{lccc}
\tabletypesize{\scriptsize}
\tablecaption{Statistics and the fraction of AGNs in 
3 morphological types of the LSBG sample (z $<$ 0.04) and the 
HSBG comparing sample.}
\tablewidth{0pt}
\tablehead{
\colhead{Sample} & \colhead{Sa} & \colhead{Sb} & \colhead{Sc}
}
\startdata
LSBGs  & 1 (14$\%$) & 6 (16$\%$) & 14 (19$\%$) \\
HSBGs  & 25 (51$\%$) & 21 (54$\%$) & 6 (33$\%$) \\
\enddata
\label{lsb_hsb_Sx}
\end{deluxetable}

\label{lastpage}
\end{document}